\documentclass{aa}  

\usepackage{graphicx}
\usepackage{txfonts}
\usepackage{lipsum}
\usepackage{subcaption}         
\usepackage{lscape}             
\usepackage{placeins}           
                                
\usepackage{multirow}
\usepackage[print-unity-mantissa=false]{siunitx}
\DeclareSIUnit{\gauss}{G} 
\DeclareSIUnit{\erg}{erg} 
\DeclareSIUnit{\dyn}{dyn} 
\DeclareSIUnit{\year}{yr} 
\DeclareSIUnit\Msun{M\textsubscript{$\odot$}}
\DeclareSIUnit{\parsec}{pc} 

\usepackage{eucal}
\usepackage[version=4]{mhchem}
\usepackage{comment}

\begin{document}




\title{Modelling mountains on accreting magnetized neutron stars}

\author{T. Brusco \inst{1}\fnmsep\inst{2}\email{tommaso.brusco@alumni.sns.it} 
    \and B. Haskell\inst{3}\fnmsep\inst{4}\email{brynmor.haskell@unimi.it}
    \and M. Razzano\inst{1}\fnmsep\inst{5}\email{MASSIMILIANO.RAZZANO@pi.infn.it}
    \and M. Bejger\inst{6}\fnmsep\inst{7}\email{bejger@camk.edu.pl}
    \and J. L. Zdunik \inst{7}\email{jlz@camk.edu.pl}
    }

\institute{Universit\`{a} di Pisa, 56127, Pisa, Italy
    \and Scuola Normale Superiore, 56126, Pisa, Italy
    \and Dipatimento di Fisica, Universit\`{a} di Milano, Via Celoria 16, 20133, Milano, Italy
    \and INFN, Sezione di Milano, Via Celoria 16, 20133, Milano, Italy
    \and INFN, Sezione di Pisa, 56127, Pisa, Italy
    \and INFN, Sezione di Ferrara, Via Saragat 1, 44122 Ferrara, Italy
    \and Nicolaus Copernicus Astronomical Center, Polish Academy of Sciences, ul. Bartycka 18, 00-716, Warsaw, Poland}


\date{Received September 30, 20XX}



\abstract{Continuous gravitational waves from accreting neutron stars in Low Mass X-ray Binaries are one of the main targets for current and next generation ground based detectors. In order to select the most promising astrophysical sources, however, reliable predictions for the signals are required, and it is therefore necessary to develop models that consistently account for the combined effects of magnetic stresses, accretion-induced heating, and the elastic response of the crust.}{We present a model for computing the quadrupolar deformation of accreting magnetized neutron stars with elastic crusts, incorporating for the first time the coupled effects of a poloidal magnetic field, deep crustal heating, and crustal elasticity.}{Perturbations to the star's structure driven by the Lorentz force density and by thermally-induced density variations are computed by solving a system of linearised deformation equations in the crust, for which we consider the full elastic response, while the ocean and core treated as barotropic fluids.}{%
  We identify a threshold accretion rate $\dot{M}_{th}$, whose value depends on crustal microphysics and the superfluid gaps in the core, above which magnetic stresses and asymmetric accretion drive deformations of opposite sign, while below this threshold their roles are reversed. The predicted eccentricities reach magnitudes up to $\varepsilon\sim 10^{-11}$, corresponding to characteristic gravitational-wave strains accessible to next-generation detectors such as the Einstein Telescope or Cosmic Explorer, but generally below the sensitivity of current LIGO, Virgo and KAGRA interferometers. These results are consistent with the non-detection of continuous gravitational waves from accreting neutron stars in Low Mass X-ray Binaries in recent observational campaigns, but highlight the need of reliable models to understand the impact of gravitational wave emission in these systems and select relevant targets for future searches.}{}

\keywords{neutron stars; gravitational waves; magnetic field; superfluid}

\maketitle

\nolinenumbers
\section{Introduction}

Continuous gravitational waves (GWs) from rotating non-axisymmetric neutron stars (NSs) are one of the main targets of current and next-generation GW detectors (\citealt{LIGOinstrument} and \citealt{2023CQGra..40r5006A}). Unlike transient signals from compact binary mergers, e.g. NS-NS binaries such as in \cite{2017PhRvL.119p1101A}, or NS -- black hole binary systems, as reported in \cite{2021ApJ...915L...5A}, these sources can be observed coherently over months or years, offering unique probes of the internal physics of NS \citep{riles}. A persistent signal requires a departure from axisymmetry, which may be produced by magnetic stresses, elastic deformations of the crust ("mountains"), free precession, or global oscillation modes such as the unstable r-modes (for a review see \citealt{GWNSreview}). Such signals have not yet been detected, but current ground based detectors now have the sensitivity to start constraining astrophysical emission scenarios \citep{haskellbejger}.

From the astrophysical perspective, one of the most interesting and often studied sources are accreting NSs in Low Mass X-ray Binaries (LMXBs). These stars are thought to be old stars that are being recycled to millisecond rotation period by accretion, and will eventually become millisecond radio pulsars \citep{1982Natur.300..728A}. 
During the accretion phase, during which these sources are detected as X-ray pulsars, one may expect many of the NSs to spin-up close to their Keplerian breakup frequency, which will be well above 1 kHz, independently of the details of the equation of state (EoS) of dense matter \citep{2007PhR...442..109L, haskellzdunik}.
Measurements of spins in these systems, however, reveal a bimodal distribution containing a population of rapidly rotating pulsars with a narrow range of frequencies $\nu_{rot} \approx \SI{575}{\hertz}$ and a maximum rotation rate of $\nu_{max}\approx 700$ Hz.
This suggests the presence of an additional mechanism that halts the spin up of the fastest rotating pulsars, a role that could be covered by a loss of energy via GW emission, due, e.g. to the presence of a mountain or unstable modes \citep{bildsten98, andersson98, Gittins}. 
While there are still uncertainties regarding the crustal EoS for a NS, the structure of this lower density region is however at least partially constrained by terrestrial nuclear physics, and it is generally established that the crust can support, without breaking, deformations, i.e. "mountains", large enough to provide a spin-down GW torque that will balance the spin-up accretion torque \citep{UCB2000, haskell06, GittinsM1, GittinsM2}

In the torque balance scenario one can therefore calculate the GW amplitude that would be needed to produce a spin-down torque that equilibrates the spin-up torque due to accretion at the observed spin frequencies of NSs in LMXBs. This is a useful benchmark for searches for continuous GW signals from such systems, which have been carried out in O3 data by the LIGO-Virgo-KAGRA collaboration \citep{2022PhRvD.105b2002A, SCO1} and, for specific cases such as Scorpius X-1, are now reaching or exceeding the torque balance limit, thus setting astrophysical constraints on the system \citep{SCOAEI, SCO2}.

In this context, where searches in LVK data can probe astrophysically realistic sectors of parameter space and constrain physical parameters, it is important to go beyond the torque balance upper limit, and predict realistic deformations of magnetised accreting NSs with elastic crusts.

While a number of authors have considered mountains due to magnetic deformations due to accretion \citep{melatosP, 2011MNRAS.417.2696P, haskell20, saurabh25} or thermal effects in the crust \citep{hutch23, hutch25, hutch26}, also in the presence of strong magnetic fields \citep{Osborne20}, no fully self consistent model that simultaneously includes accretion physics, elasticity and magnetic field exists. All these effects are however required, as they contribute to generating and maintaining the quadrupolar deformations that give rise to continuous GW emission, and their interplay, as we will see, can lead to subtle cancellations and changes in overall shape of the star.

The model presented here therefore takes a first step towards self-consistently computing the quadrupolar deformation of a NS consisting of an elastic crust and a fluid core, endowed with a poloidal magnetic field, of intensity $B_p$ at the pole at the surface, and accreting at a total rate $\dot{M}$. Specifically we calcule equilibria around a spherical background in hydrostatic and thermal equilibrium, and consider the effect of the magnetic field, deep crustal heating reactions due to accretion and elasticity as linearised perturbations. 

In Sec.~\ref{back} we therefore present first the background model for our star and discuss the problem setup, in Sec.~\ref{sec:def} we outline our procedure to solve the linearised perturbation problem and obtain the deformations of the NS and in Sec.~\ref{results} we present and discuss our results. Finally, conclusions are drawn in Sec.~\ref{conclu}, pointing out 
the peculiar interplay of accretion rate and accretion asymmetry obtained through the model and the possibility of confirmation through observations by future GW detectors.




\section{The background model}
\label{back}
The background model of the star describes the unperturbed, spherically symmetric NS. 
The accreted BSk21 model presented in~\cite{EoS18,EoS22} is used for the EoS 
and the composition of the star. In addition, a total mass of $M=\SI{1.4}{\Msun}$ is assumed.

\subsection{Background density}
The radial density profile of the unperturbed star is an approximation of the one obtained 
by solving the Tolman-Oppenheimer-Volkoff (TOV) equation, first derived in~\cite{TOV}, 
for a total mass of $M=\SI{1.4}{\Msun}$ and with accreted BSk21 EoS presented in~\cite{EoS22}.

The crust-core interface sits at the value $r_{cc}$ of the radial coordinate where 
the TOV density profile matches the crust-core density 
$\rho_{cc}=\SI{1.34e14}{\gram\per\centi\meter\cubed}$ defined by the BSk21 EoS. 
That value is $r_{cc} = \SI{11.55}{\kilo\meter}$.

A total radius of $R=\SI{12.66}{\kilo\meter}$ is also found for the star by solving the TOV equation.

In the core ($r<r_{cc}$), the TOV profile is approximated with 
\begin{equation}\label{eq:rhoCore}
    \rho (r) = \rho_0 \eta \frac{\sin(\eta r)}{r},
\end{equation}
while the equation of state in the same region is approximated with a polytropic one with 
adiabatic index $\gamma=2$. 
In particular, the central density $\rho_0$ in equation~\ref{eq:rhoCore} is found through a least-squares fitting to the numerical solution of the TOV equation with the constraint that the $\eta$ 
parameter is such that $\rho (r_{cc}) = \rho_{cc}$.

On the other hand, the approximated background density profile in the crust is
\begin{equation}\label{eq:rhoCrust}
    \rho (r) = \rho_{cc} \left(1-\frac{r-r_{cc}}{R_0 -r_{cc}}\right)^7,
\end{equation}
where $R_0$ is the effective radius of the star, which corresponds to the radial coordinate at 
which the approximated density is 0. 
The value of $R_0$ is found through a least square fitting of the profile in Eq.~\ref{eq:rhoCrust} to the numerical solution of the TOV equation.

\subsection{Thermal structure}
\label{sec:bgTemp}

The thermal structure of the star accounts for the subdivision of its outer regions in 
several layers. 

The outmost layer is the ocean, whose outer boundary is the \ce{H}/\ce{He} burning layer 
at $r_{\ce{H}/\ce{He}} = \SI{12.65}{\kilo\meter}$, 
which corresponds to the point where the background density 
is $\rho=\SI{1e7}{\gram\per\centi\meter\cubed}$. This is the highest density at which 
the \ce{H}/\ce{He} burning of accreting matter happens, according to~\cite{HHeburn}. The burning 
reactions are assumed to leave only \ce{^{56}Fe} ashes. The temperature inside the ocean is 
assumed to be constant in the ocean. Its value depends on the accretion rate, 
according to~\cite{schatz}, as
\begin{equation}\label{eq:surfaceT}
    T_S = \SI{1.13e8}{\kelvin} \left( \frac{\dot{M}}{\SI{1e-10}{\Msun\per\year}} \right)^{\frac{2}{7}}.
\end{equation}
The inner boundary of the ocean is the radius $r_S$ where the ratio of Coulomb energy to thermal energy is  
\begin{equation}
\Gamma_{Coul} := \frac{Z^2 q_e^2}{k_B T_S} \left( \frac{4\pi \rho(r_S)}{3 m_u} \right)^{\frac{1}{3}} = 175,
\end{equation}
where $Z=26$ is the atomic number of elements in the ocean, $q_e$ is the electron elementary charge, 
$k_B$ is Boltzmann's constant and $m_u$ is the atomic mass unit. This is the usual criterion for 
the melting of the crust, used e.g. in~\cite{HaenselBook} and~\cite{hutch23}.

In the one-component plasma (OCP) assumption, the solid crust is divided into several layers, 
each of which has a dominant element, defined by the accreted BSk21 model presented in~\cite{EoS18}. 
More realistically, in the inner region of each layer, the concentration of the dominant element 
diminishes as it is converted into the main component of the next layer by electron 
capture reactions in the form \ce{(Z_1^{(i)}, A_1^{(i)}) + n e -> (Z_2^{(i)}, A_2^{(i)}) + n \nu_e} \citep{lami}.

In the $i$-th layer the mass fraction $X_i$ of the dominant element \ce{(Z_1^{(i)}, A_1^{(i)})} 
is found consistently with the unperturbed temperature profile $T_i$ and heat flux $F_i$ in the 
same region. 
The equations for that in each layer, already presented in~\cite{UCB2000}, are
\begin{subequations}\label{eq:thermStruct}
\begin{eqnarray}
    \frac{\partial F_i}{\partial r} &=& -\frac{2}{r} F_i + \mathcal{H}_S^{(i)} (\dot{M}) + \mathcal{H}_D^{(i)} (r, T_i, X_i) - \mathcal{H}_\nu^{(i)} (r, T_i, X_i)\,,\\
    \frac{\partial T_i}{\partial r} &=& - \frac{F_i}{K_i (r, T_i, X_i)}\,,\\
    \frac{\partial X_i}{\partial r} &=& \frac{4 \pi r^2}{\dot{M}} X_i \rho R_{ec}^{(i)} (r, T_i, X_i)\,,
\end{eqnarray}
\end{subequations}
where $\mathcal{H}_S^{(i)}$ is the shallow crustal heating (SCH), which is the heating released by 
accretion in the top layers of the crust, $\mathcal{H}_D^{(i)}$ is the deep crustal heating (DCH), 
which is the heating released in all layers of the crust by reactions, $\mathcal{H}_\nu^{(i)}$ is the 
neutrino cooling, $K_i$ is the thermal conductivity and $R_{ec}^{(i)}$ is the 
reaction rate of the electron capture that happens in the layer. 

SCH is assumed to be spread across the three outermost layers of the crust. It is given by
\begin{equation}
    \mathcal{H}_S^{(i)} = 
        \begin{cases}
            \frac{3\dot{M} Q_S}{{4\pi} m_u \left(r_S^3-r_3^3 \right)} & \quad \text{if } i\leq 3 \\
            0 & \quad \text{otherwise},
        \end{cases}
\end{equation}
where $r_3$ is the inner boundary of the third layer from the surface and $Q_S=\SI{1.5}{\mega\eV}$ 
is the heat released per accreted nucleon by SCH \citep{hutch23}.

The expressions for DCH and electron capture reaction rates are taken from~\cite{UCB2000}, 
in order to capture their dependency on the mass fraction of reactant $X_i$. The reactions 
considered are those presented in~\cite{EoS18} for the accreted BSk21 model (see \cite{Gusakov} for a discussion of how neutron diffusion may impact these results). In particular, 
the DCH in the $i$-th layers is 
\begin{equation}
    \mathcal{H}_D^{(i)} = \frac{Q_i}{m_u} \frac{X_i \rho}{1-X_{n,1}^{(i)}} R_{ec}^{(i)} (r, T_i, X_i),
\end{equation}
where $Q_i$ is the heat released per accreted nucleon by the reaction that happens in the $i$-th 
layer and $X_{n,1}^{(i)}$ is the mass fraction of free neutrons at the outer boundary of that layer, 
both of which values are listed in~\cite{EoS18}. 

The reaction rate of the electron capture reaction in the $i$-th layer is
\begin{equation}
    R_{ec}^{(i)} = \frac{2 \ln 2}{ft} \frac{\left(E_{th}^{(i)} \right)^2 \left(k_B T_i \right)^3}{(m_e c^2)^5} e^{\frac{E_F^{(i)} (r, X_i) -E_{th}^{(i)}}{k_B T_i}},
\end{equation}
where $ft \approx \SI{1e4}{\second}$ as the $ft$ parameter for all the reactions, $E_{th}^{(i)}$ 
is the threshold energy of the electron capture in the $i$-th layer, as listed in~\cite{EoS18} 
and $m_e$ is the mass of an electron. The electron Fermi energy profile across the $i$-th layer 
is indicated by $E_F^{(i)}$, which can be expressed as
\begin{equation}
    E_F^{(i)} = c \hbar \left( \frac{3\pi^2}{m_u} \frac{\rho (r)}{\mu_e^{(i)} (X_i)} \right)^{\frac{1}{3}},
\end{equation}
where $\mu_e$ is the mean molecular weight per electron in the $i$-th layer, 
which can be derived from $X_i$. Only the electron capture reaction rate is considered since, 
in case a pycnonuclear reaction happens in a layer, it is assumed to always take place right after 
the electron capture.

The mean molecular weight per electron in the $i$-th layer \citep{UCB2000} depends on the mass fraction of the main component as
\begin{equation}
    \frac{1}{\mu_e^{(i)}} = \left[ \frac{Z_1^{(i)}}{A_1^{(i)}} - \frac{Z_2^{(i)}}{A_2^{(i)}} \left( 1 - \frac{X_{n,1}^{(i)}-X_{n,2}^{(i)}}{1-X_{n,2}^{(i)}} \right) \right] X_i + \left(1-X_{n,2}^{(i)} \right) \frac{Z_2^{(i)}}{A_2^{(i)}}
\end{equation},
where $X_{n,2}^{(i)}$ is the mass fraction of free neutrons at the inner boundary of the $i$-th layer.

For the neutrino cooling, only bremsstrahlung of electrons on nuclei is taken into account. In that 
case, for each layer, it is expressed in~\cite{1996AstL...22..491Y} as
\begin{equation}
    \mathcal{H}_\nu^{(i)} = H_\nu \Lambda_{br} \frac{\rho}{\SI{1e12}{\gram\per\centi\meter\cubed}} \left( \frac{T_i}{\SI{1e8}{\kelvin}} \right)^6 \frac{A_{cell}^{(i)}}{\left(\mu_e^{(i)} \right)^2} \left(1-X_n^{(i)} \right),
\end{equation}
with $H_\nu = \SI{3.229e11}{\erg\per\second\per\centi\meter\cubed}$ and 
where $\Lambda_{br} \approx 1$ is the Coulomb logarithm of the bremsstrahlung reaction, 
while $A_{cell}^{(i)}$ is the total number of nucleons per Wigner-Seitz cell of the lattice and
$X_n^{(i)}$ is the mass fraction of neutrons, both of which values can be determined from $X_i$ 
at each point of the layer. In particular, $A_{cell}^{(i)}$ changes only across layers that include 
a pycnonuclear reaction, in which case its expression is
\begin{equation}
    A_{cell}^{(i)} = 2 A_{cell, 1}^{(i)} \frac{1- X_{n,1}^{(i)}}{1-X_{n,1}^{(i)}+X_i},
\end{equation}
where $A_{cell,1}^{(i)}$ is the total number of nucleons per cell at the outer boundary of the 
$i$-th layer.
On the other hand the mass fraction of free neutrons is
\begin{equation}
    X_n^{(i)} = \frac{X_{n,1}^{(i)} - X_{n,2}^{(i)}}{1-X_{n,1}^{(i)}} X_i + X_{n,2}^{(i)},
\end{equation}
using the formula from~\cite{UCB2000}.

\begin{figure*}[ht]
    \centering
    \includegraphics[scale=.5]{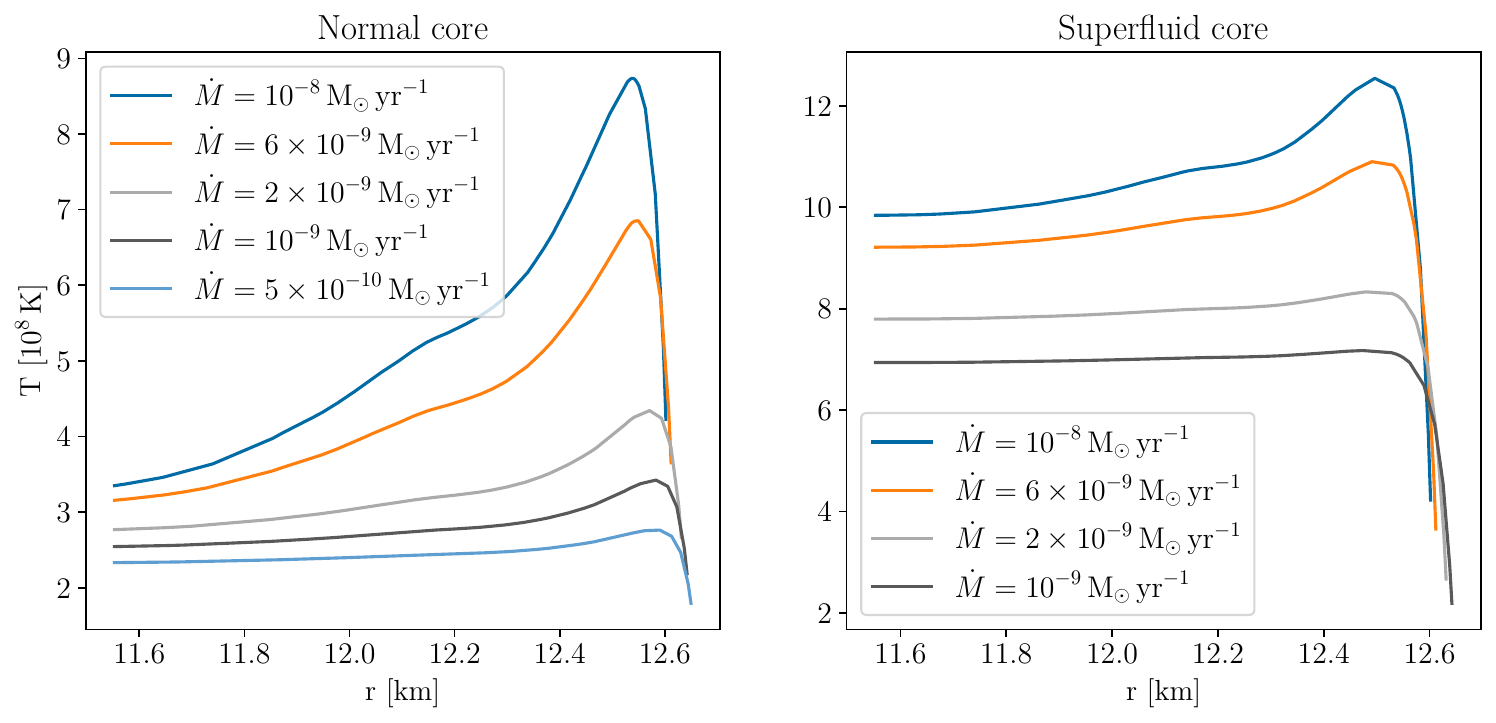}
    {\caption{Temperature profiles of unperturbed stars with normal and superfluid cores. Left panel: stars with a 
    superfluid energy gap of $\Delta=0$ in the core. Right panel: superfluid core stars with $\Delta = \SI{1}{\mega\eV}$.}
    \label{fig:bgTemp}}
\end{figure*}

The thermal conductivity inside the crust, according to~\cite{ziman}, is
\begin{equation}
    K_i = \frac{k_B^2 c}{\hbar} \left( \frac{\pi}{9 m_u} \frac{\rho (r)}{\mu_e^{(i)} (X_i)} \right)^{\frac{2}{3}} T_i \, t_{rel}^{(i)} (r, T_i, X_i)
\end{equation},
where $t_{rel}^{(i)}=\frac{1}{\nu_{ep}^{(i)} (T_i) + \nu_{eQ}^{(i)} (r, X_i)}$ is the collisional relaxation time, 
with $\nu_{ep}$ and $\nu_{eQ}$ rates of the scattering processes considered for the conduction 
of heat in the crust, respectively the electron-phonon scattering and the electron-impurity 
scattering. The expressions of both rates are taken from~\cite{brown09}. 
In particular, for electron-phonon scattering, it is
\begin{equation}
    \nu_{ep}^{(i)} = \SI{1.25e18}{\hertz} \frac{T_i}{\SI{1e8}{\kelvin}}.
\end{equation}
On the other hand, the rate of electron-impurity scattering is
\begin{equation}
    \nu_{eQ}^{(i)} = \SI{1.77e18}{\hertz} \, Q_{imp} \, \Lambda_{imp} \left(\mu_e^{(i)} (X_i) \right)^{\frac{2}{3}} \frac{\rho(r)^{\frac{1}{3}}}{A_{cell}^{(i)} (X_i)}, 
\end{equation}
where $\Lambda_{imp} \approx 1$ is the assumed Coulomb logarithm of the scattering process, 
while $Q_{imp}$ is called the impurity factor, which accounts for elements present in the crust 
beyond the OCP assumption. It is defined as 
\begin{equation}
    Q_{imp} := \frac{1}{n_{ion}} \sum_j n_j \left( Z_j - \left< Z \right> \right)^2,
\end{equation}
where $n_{ion}$ is the number density of atomic nuclei, $\left< Z \right>$ is the average 
atomic number of elements in the crust, the sum is performed across all atomic species and $n_j$ and 
$Z_j$ are respectively the number density and the atomic number of the species $j$. It is assumed 
that $Q_{imp} \approx 1$ according to the results of \cite{ootes19}, \cite{parikh19}, \cite{cumming17}, \cite{merritt16}, \cite{turlione15} and \cite{brown09}.  

Equations~\ref{eq:thermStruct} are solved for each layer with the following boundary conditions.
The outer boundary of the outermost crustal layer is at $r_S$, the inner boundary of the 
innermost layer is at $r_{cc}$ and the inner boundary of each other layer sits 
where $X_i =\num{2e-6}$. 
Both the temperature and the heat flux are assumed to be continuous between a layer and the next 
one, while $X_i=1$ at the outer boundary of each layer. Additional boundary conditions are that the 
temperature at the outer boundary of the outermost layer is $T_S$
and that the following condition is satisfied at the crust core interface
\begin{equation}
    L_{Urca} (T_{cc}, \rho_{cc}) + 4\pi r_{cc}^2 F_{cc} = 0,
\end{equation}
where $T_{cc}$ and $F_{cc}$ are the temperature and the heat flux at the crust-core interface, while 
$L_{Urca}$ is the Urca luminosity of the core, defined in~\cite{TeukolskyBook} as
\begin{equation}
    L_{Urca} = \SI{5.31e31}{\erg\per\second} \frac{M}{\si{\Msun}} \left( \frac{\rho_{sat}}{\rho} \right)^{\frac{1}{3}} \left( \frac{T}{\SI{1e8}{\kelvin}} \right)^8 e^{-\frac{\Delta}{k_B T}},
\end{equation}
with $\rho_{sat}=\SI{2.8e14}{\gram\per\centi\meter\cubed}$, according to~\cite{HaenselBook}, being 
the nuclear saturation density for large atomic nuclei and $\Delta$ the superfluid energy gap 
in the star's core.

Equations~\ref{eq:thermStruct} are solved considering either a superconducting core 
with $\Delta = \SI{1}{\mega\eV}$ or a normal core with $\Delta=0$. In the first case they are 
solved for the following values of the accretion rate: $\dot{M} = \SI{1e-8}{\Msun\per\year}$, 
$\dot{M} = \SI{6e-9}{\Msun\per\year}$, $\dot{M} = \SI{2e-9}{\Msun\per\year}$ and 
$\dot{M} = \SI{1e-9}{\Msun\per\year}$. The same values are considered in the case of a normal core 
with the addition of $\dot{M} = \SI{5e-10}{\Msun\per\year}$. The resulting background temperature 
profiles are shown in Fig.~\ref{fig:bgTemp}. 

The temperature profile $T$ shown in Fig.~\ref{fig:bgTemp} is obtained by juxtaposing the 
profiles $T_i$ of each layer. Similarly, profiles across the whole crust can be obtained for the 
thermal conductivity $K$, the mass fraction of free neutrons $X_n$, the mean molecular weight per 
electron $\mu_e$, the electron Fermi energy $E_F$ and the collisional relaxation time $t_{rel}$, 
which are all relevant quantities when determining the deformations of the star.

\subsection{Magnetic field}
\label{sec:magn}

The magnetic field of the star is assumed to be poloidal only and to connect to an 
external dipole field without current sheets at the surface. 
Indeed the toroidal component would be negligible in the crust, where the most relevant deformation 
is produced.

With those assumptions, the magnetic field can be parametrized as
\begin{equation}\label{eq:poloidalB}
    \vec{B} = \frac{B_p R_0^2}{2} \left[ \frac{2}{r^2} \mathcal{A} \cos\theta \hat{r}- \frac{1}{r} \frac{\partial \mathcal{A}}{\partial r} \sin\theta \hat{\theta}\right],
\end{equation}
where $B_p$ is the magnetic field at the poles of the star and  $\mathcal{A}(r)$ 
is a function of the radial coordinate hereafter called the magnetic function.

The magnetic function, and thence the shape of the magnetic field, is found in the core through the Grad-Shafranov 
equation
\begin{equation}\label{eq:GScore}
    \vec{\nabla} \times \left( \frac{\vec{f}}{\rho}\right)= 0,
\end{equation}
where $\vec{f}=\frac{1}{4\pi} \left(\vec{\nabla} \times \vec{B}\right) \times \vec{B}$ is the Lorentz force density.

On the other hand, it is not possible to apply the Grad-Shafranov equation in the solid crust, since elastic forces 
are present there in addition to the magnetic force. However, with the additional assumption that the magnetic field 
is static in the crust, considering a very large electrical conductivity, a similar equation can be found:
\begin{equation}\label{eq:GScrust}
    \vec{\nabla} \times \left( \frac{A_{cell}}{Z \rho} \vec{f}\right) = 0.
\end{equation}

The analytical solution of Eqs.~\ref{eq:GScore} and~\ref{eq:GScrust} was found in~\cite{Gourg13}. 
The two solutions are connected assuming the regularity of the magnetic field at the center 
and its continuity (without current sheets) at the crust-core interface. It is also assumed that  
the component along $\hat{\theta}$ of the Lorentz force density is continuous at the crust-core 
interface, otherwise the magnetic force would push free charges in the radial direction, 
creating a flow across that surface. Thus the expression of the magnetic function across all the star 
is found and so all components of the Lorentz force density are known.

\section{Deformations of NSs}
\label{sec:def}

The sources of the deformations considered in the NS are its magnetic field and the 
temperature deformation that it induces. Both are assumed to be perturbative. 

The perturbative part of all vector quantities is decomposed in vector spherical harmonics (VSH),
for which the same convention as in~\cite{barreraVSH} is used. The $z$-axis of the VSH is 
the magnetic axis, rather than the rotation axis, of the star and only multipoles with $\ell=1$ are 
considered since they are the only ones that matter for the emission of gravitational waves.

\subsection{Deformations in the core}
\label{sec:core}

The core of the NS is assumed to be liquid in the sense that there are no elastic forces in its 
interior. In this context, the perturbed conservation of momentum equation is
\begin{equation}\label{eq:momLiquid}
    \vec{\nabla} \delta p = \vec{f} - \delta \rho \vec{\nabla} \Phi,
\end{equation}
where $\delta p$ and $\delta \rho$ are the pressure and density perturbations, respectively, and 
$\Phi$ is the gravitational potential. The density perturbation that deforms the core solves this 
equation since it is the equilibrium configuration in the presence of a magnetic field that 
corresponds to the Lorentz force density $\vec{f}$.

In addition, a barotropic EoS is assumed in the core, which means that
\begin{equation}
    \delta p = c_S^2 \, \delta \rho,
\end{equation}
where $c_S^2 := \frac{\partial p}{\partial \rho}$ is the speed of sound in the core.

In the assumption that
\begin{equation}\label{eq:cs2CoreCond}
    \frac{\partial c_S^2}{\partial r} \simeq - \frac{\partial \Phi}{\partial r},
\end{equation}
which holds exactly when the EoS of the core is approximated to a polytropic with adiabatic 
index $\gamma =2$, Eq.~\ref{eq:momLiquid}, decomposed in VSH, reduces to
\begin{equation}\label{eq:defCore}
    \delta \rho_{\ell m} = \frac{r}{c_S^2} f^{(\bot)}_{20}.
\end{equation}

Finally, it should be noted that Eq.~\ref{eq:defCore} also holds in the ocean, which is 
also assumed to be liquid and to be described by a barotropic EoS that approximately satisfies Eq.~\ref{eq:cs2CoreCond}.

\subsection{Perturbed flux equation}
\label{sec:therm}

In general, the heat flux is affected by the magnetic field, since heat conductors are electrons.
The full formula, from~\cite{Urpin80}, is 
\begin{equation}\label{eq:magnHeat}
    \vec{F} = - K \left[ \vec{\nabla} T + \frac{x_m^2}{\varpi} \vec{B} \times(\vec{B} \times \vec{\nabla} T) + \frac{x_m B_p}{\varpi} \vec{B} \times \vec{\nabla} T \right],
\end{equation}
where $\varpi := B_p^2 + \left(x_m |\vec{B}|\right)^2$ and $x_m := \frac{q_e c}{E_F} B_p t_{rel}$, 
which is called the magnetization parameter, as in~\cite{Osborne20}. That parameter corresponds to 
the ratio between the gyromagnetic frequency in the star and the 
collisional relaxation frequency and accounts for the magnitude of the magnetic field inside the crust 
with respect to thermal conduction. In particular, the magnetic field is considered perturbative if 
at all points in the crust $x_m<1$.

On the other hand, the heating equation $\vec{\nabla} \cdot \vec{F} = \mathcal{H}$, 
with $\mathcal{H}$ the total heating (including DCH, SCH and neutrino cooling), is not affected by 
the presence of a magnetic field.

In principle, background quantities such as conduction and heating depend on density, 
temperature and composition (parametrized by $\mu_e$). However, it is assumed that the variations 
of those quantities under small variations of temperature are more relevant than their variations 
under density or composition perturbations. In other words the variation of a 
quantity $\psi (\rho, T, \mu_e)$ is
\begin{equation}
    \frac{\delta \psi}{\psi} =  \left.\frac{\partial \ln \psi}{\partial \ln T} \right\vert_{\rho, \mu_e} \frac{\delta T}{T}.
\end{equation}
This also accounts for the assumption, already in~\cite{UCB2000}, 
that the composition changes only because of the rigid displacement of the one component layers. 

In addition, perturbing equation~\ref{eq:magnHeat}, requires special attention. First of all, the final perturbed equation should include only VSH components with $\ell = 2$ and $m=0$, since those are the only ones that will be relevant in determining the eccentricity of the star. For this reason, when imposing a purely poloidal magnetic field, the last term on the right hand side of equation~\ref{eq:magnHeat} can be neglected, since $\vec{B} \times \vec{\nabla} T$ is along the $\hat{\phi}$ direction for a poloidal field, and such a vector field would vanish when projectd along $\ell=2, \ m=0$ VSH. Considering this, equation~\ref{eq:magnHeat} is perturbed taking $T \mapsto T(r) + \delta T_{20} Y_{20}$ and $K \mapsto K(r) + \delta K_{20} Y_{20}$, and excluding the $-K \frac{\partial T}{\partial r}$ term on the right hand side (which was already included in the background equations) and all terms containing both $\delta T_{20}$ and $\delta K_{20}$. This perturbation includes the assumption that the variation of $t_{rel}$ under a temperature perturbation can be neglected. Then the equation is projected along the $\ell=2, \ m=0$ VSH, and finally it is expanded for low values of the magnetic field, keeping terms up to second order in $x_m$.

With these premises and considering the poloidal only magnetic field of Eqs.~\ref{eq:poloidalB}, and equation~\ref{eq:magnHeat} gives
\begin{eqnarray} \label{eq:deltaT}
    \frac{\partial \delta T_{20}}{\partial r} &=& S_r x_m^2 - \frac{1+ \mathcal{B}_1 x_m^2}{K} \delta F^{(r)}_{20} - \left( f_K \frac{\partial T}{\partial r} + \mathcal{B}_2 x_m^2 \right) \delta T_{20}\\
   \delta F^{(\bot)}_{20} &=& \frac{r}{\lambda} S_\bot x_m^2 + \frac{r}{\lambda} \mathcal{B}_2 x_m^2 \delta F^{(r)}_{20} - \left( 1- \frac{r}{\lambda} \mathcal{B}_3 \right) K \delta T_{20},
\end{eqnarray}
where 
\begin{subequations}
\begin{eqnarray}
    \lambda &:=& \ell (\ell +1) = 6 \\
    f_K  &=& \frac{K}{T} \left( 1- t_{rel} \nu_{ep} \right) \\
    \mathcal{B}_1 &:=& \frac{5}{42} R_0^4 \left( \frac{1}{r} \frac{\partial \mathcal{A}}{\partial r} \right)^2 \\
    \mathcal{B}_2 &:=& \frac{R_0^4}{7} \frac{\mathcal{A}}{r^4} \frac{\partial \mathcal{A}}{\partial r} \\
    \mathcal{B}_3 &:=& \frac{72}{7} R_0^4 \frac{\mathcal{A}^2}{r^6} \\
    S_r &:=& - \frac{1}{3} \sqrt{\frac{\pi}{5}} R_0^4 \left( \frac{1}{r} \frac{\partial \mathcal{A}}{\partial r} \right)^2 \frac{\partial T}{\partial r} \\
    S_\bot &:=& -2 \sqrt{\frac{\pi}{5}} R_0^4 \frac{\mathcal{A}}{r^4} \frac{\partial \mathcal{A}}{\partial r} \frac{\partial T}{\partial r} K.
\end{eqnarray}
\end{subequations}

On the other hand, perturbing the heat equation gives
\begin{equation}\label{eq:deltaF}
    \frac{\partial \delta F^{(r)}_{20}}{\partial r} = \frac{\lambda}{r} \delta F_{20}^{(\bot)} + f_{\mathcal{H}} \delta T_{20} - \frac{2}{r} \delta F^{(r)}_{20}\,,
\end{equation}
where
\begin{equation}
    f_{\mathcal{H}} = \frac{1}{T} \left[ 6 \mathcal{H}_\nu + \sum_i \left( 3- \frac{E_F^{(i)}-E_{th}^{(i)}}{k_B T_i} \right) \mathcal{H}_D^{(i)} \right]\,,
\end{equation}
with the sum in the last term performed over the single component layers.


Equations~\ref{eq:deltaT} and~\ref{eq:deltaF} form a non-homogeneous system of two linear 
differential equations. The source term for this set of equation is proportional to $x_m$ and represents an indirect inluence of the magnetic field on the temperature perturbation and thence in the deformations of the NS.

In order to solve the system of equations~\ref{eq:deltaT} and~\ref{eq:deltaF}, the homogeneous part is solved numerically twice, each time setting 
a different independent variable to be equal to 1 at one end of the integration interval. The 
non homogeneous system is instead solved by setting to 0 all variables at one end of the 
integration interval. The three solutions are then combined linearly with appropriate coefficients, such that they satisfy the boundary conditions. Numerical solutions of systems of differential equations 

Equations~\ref{eq:deltaT} and~\ref{eq:deltaF} are solved together between $r_{cc}$ and $r_S$ in 
order to find the temperature perturbation profile $\delta T_{20}$ in the crust. The boundary 
condition at the crust-core interface for solving the equations is $\delta T_{20} (r_{cc}) =0$ since 
the thermal conductivity is much larger in the core than in the crust, as assumed also 
by~\cite{UCB2000} and~\cite{Osborne20}. At the crust-ocean interface, instead, the temperature 
perturbation is determined by accreted matter depositing on the crust. Indeed, as pointed out 
in~\cite{haskell20}, the accretion flow is channeled along magnetic field lines and falls onto 
the polar caps of the NS, which are the regions, close to the magnetic poles, where the star's 
surface is crossed by magnetic field lines that are open at the Alfvén radius $r_A$. In this 
framework, $\dot{M}$ is the accretion rate averaged across the star's surface, but the local 
accretion rate is zero outside polar caps. This asymmetric
accretion produces an asymmetry in the temperature at the outer boundary of the ocean (Eq.~\ref{eq:surfaceT}). However, the 
temperature asymmetry at the outer boundary of the crust may be dampened with respect to the one 
at the star's surface, either because accreted matter spreads across the surface in the width of 
the ocean or because the high thermal conductivity of that region flattens perturbations. All this 
being considered, the boundary condition for the temperature perturbation at the crust-ocean interface is 
\begin{equation}
    \delta T_{20} (r_S) = \frac{\sqrt{5\pi}}{7} \left( a + a^2 \right) T_S
\end{equation}
with
\begin{equation}
    a := D \sqrt{1- \sqrt{\frac{R}{r_A}}},
\end{equation}
where $D$ is a free parameter of the model, hereafter called \emph{accretion asymmetry}, 
representing the spreading of accreted matter in the ocean. 
If $D=1$ the angular size of the region of the crust that is heated by accretion is 
the same as that of the polar caps, while for $D=0$ the whole crust is equally heated by 
accretion, and there is no temperature asymmetry.
The surface temperature perturbation described here has a $\ell=2$, $m=0$ symmetry, which 
justifies the simplifying assumption of considering only the $\delta T_{20}$ component of the 
temperature perturbation.

To determine $D$ consistently, it would be necessary to also solve the MHD equilibrium equations 
in the ocean of the star, and match the solutions at the boundary. As this is beyond the scope of 
the current work, a fiducial range of values will be taken for that parameter while solving the deformation equations in Sec.~\ref{results}.


The Alfvén radius of the NS is determined as in~\cite{1977ApJ...215..897E}:
\begin{equation}
    \begin{split}
        r_A &=& \SI{3.53e3}{\kilo\meter} \left( \frac{B_p}{\SI{1e12}{\gauss}} \right)^{\frac{4}{7}} \left( \frac{R}{\SI{10}{\kilo\meter}} \right)^{\frac{12}{7}} \\
        && \times \left( \frac{\dot{M}}{\SI{1e-9}{\Msun\per\year}} \right)^{-\frac{2}{7}} \left( \frac{M}{\SI{1.4}{\Msun}} \right)^{-\frac{1}{7}}.
    \end{split}
\end{equation}

\begin{figure*}[ht]
    \centering
    \includegraphics[scale=.5]{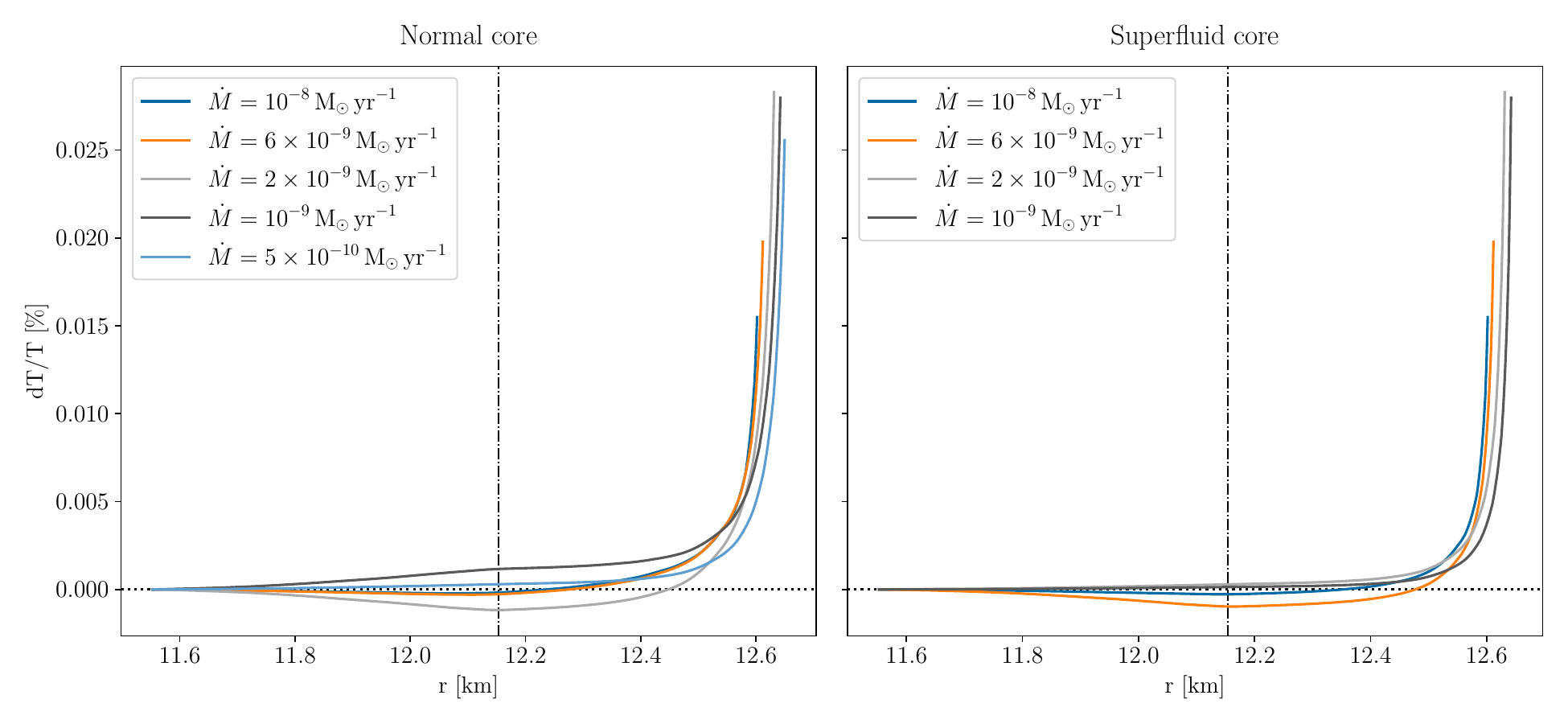}
    {\caption{Various temperature perturbation profiles. For all profiles shown $B_p=\SI{1e8}{\gauss}$ and $D=0.001$, while the kind of core and the total accretion change. The left panel considers 
    $\Delta=0$, while in the right panel $\Delta =\SI{1}{\mega\eV}$. In each panel, 
    the dash-dotted line marks the center of the capture layer whose reaction releases the 
    most energy per accreted nucleon, in the BSk21 model.}
    \label{fig:deltaT}}
\end{figure*}

Figure~\ref{fig:deltaT} shows some examples of the temperature perturbation profiles thus obtained.

\subsection{Deformations of the crust}
\label{sec:defCrust}

The crust is assumed to be solid and composed of an atomic nuclei lattice. In this case, internal 
elastic forces should be considered and the perturbed conservation of momentum equation is
\begin{equation}\label{eq:mom}
    \vec{\nabla} \cdot \pmb{\tau} = -\vec{f} + \delta \rho \vec{\nabla} \Phi,
\end{equation}
where $\boldsymbol{\tau}$ is the full stress tensor, which represents the force per unit area that 
deforms a volume element of the crust in each direction.

Following the approach of~\cite{UCB2000}, the full stress tensor is expressed as
\begin{equation}\label{eq:stress}
    \pmb{\tau} = -\delta p \pmb{g} + \pmb{\sigma},
\end{equation}
where $\boldsymbol{g}$ is the metric tensor of the flat 3D metric and $\boldsymbol{\sigma}$ is 
the strain tensor, which quantifies elastic forces in the crust. 
In components, the strain tensor is
\begin{equation}
    \sigma_{ij} = \mu \left[ \nabla_i \xi_j + \nabla_j \xi_i - \frac{2}{3} g_{ij} \vec{\nabla}\cdot\vec{\xi} \right],
\end{equation}
where $\vec{\xi}$ is the displacement vector field, which is the displacement of volume elements 
of the crust at each point, and $\mu$ is the shear modulus, which accounts for the rigidity of 
the crust. According to~\cite{haskell06} and~\cite{ogata90}, the expression of the shear modulus is
\begin{equation}
    \mu = \SI{1.28e31}{\erg\per\centi\meter\cubed} \frac{Z^2}{A_{cell}^{\frac{4}{3}}} \left( \frac{\rho}{\rho_{cc}} \right)^{\frac{4}{3}}.
\end{equation}

In order to capture thermal deformations, $\delta p$ must be expressed in terms of both 
the displacement and the temperature variation, and a non-barotropic EoS is needed for that. 
In particular, the EoS of the crust is approximated with
\begin{equation}\label{eq:pressCrust}
    p = p_{n, 0} \left( \frac{X_n \rho}{X_{n, cc} \rho_{cc}} \right)^{\frac{5}{3}} + \frac{\hbar c}{4} \left( \frac{3 \pi^2}{m_u^4} \right)^{\frac{1}{3}} \left( \frac{\rho}{\mu_e} \right)^{\frac{4}{3}},
\end{equation}
where $X_{n,cc}$ is the mass fraction of free neutrons at the crust-core interface and
$p_{n,0}$ is a constant with the dimensions of a pressure and chosen such that the unperturbed 
pressure is continuous at the crust-core interface. Equation~\ref{eq:pressCrust} implies that 
the total pressure inside the crust is only due to the degeneracy pressures of the Fermi gases 
of free electrons, which are assumed to be relativistic, and free neutrons, which are assumed to be 
non-relativistic and non-interacting.

In Eq.~\ref{eq:pressCrust}, the pressure in the crust depends on the temperature 
through $\mu_e$. Indeed, a temperature variation changes the depth at which each electron 
capture reaction happens, which changes the crust's composition, thence $\mu_e$. 
As in~\cite{UCB2000}, 
layers are assumed to be displaced rigidly, without changing their width, and the center of each 
reaction layer is defined as the point where the local compression time scale equals the 
inverse of the electron capture rate. Since this last quantity depends on $T$, the 
pressure $p_c^{(i)}$ at the center of the $i$-th layer also depends on temperature. The radius 
corresponding to the center of the $i$-th capture layer is indicated with $r_c^{(i)}$.

To sum up, the displacement of the $i$-th reaction layer under a temperature 
perturbation $\delta T$ is
\begin{equation}
    \Delta z_i = - \left( \left.\frac{\mathrm{d} p}{\mathrm{d} r} \right|_{r=r_c^{(i)}} \right)^{-1} \frac{\partial p_c^{(i)}}{\partial T} \, \delta T =: g_z^{(i)} \frac{\delta T}{T}.
\end{equation}

Since the composition changes under temperature perturbation only through the radial displacement 
of layers, the Lagrangian perturbation of the mean molecular weight per electron in the $i$-th 
layer is
\begin{equation}
    \Delta \mu_e^{(i)} = - \frac{\partial \mu_e^{(i)}}{\partial r} \Delta z_i.
\end{equation}

Then, the Eulerian pressure perturbation can be expressed 
in terms of the density and temperature variations as 
\begin{equation}\label{eq:deltap}
    \delta p = c_S^2 \, \delta \rho - q_T \, \delta T,
\end{equation}
with
\begin{subequations}
\begin{eqnarray}
    c_S^2 &=& \left.\frac{\partial p}{\partial \rho}\right|_{\mu_e} \\
    q_T&:=& \left.\frac{\partial p}{\partial \mu_e}\right|_{\rho} \frac{\partial \mu_e}{\partial r} \frac{g_Z}{T}.
\end{eqnarray}
\end{subequations}
It should be noted that in the crust the speed of sound $c_S^2$ must be computed at 
constant composition.

The deformations of the crust are more conveniently described in terms of the displacement and the 
traction vector, which is defined as $\vec{t} := \hat{r} \cdot \pmb{\tau}$. 
In this formalism, using Eqs.~\ref{eq:stress} and~\ref{eq:deltap} and decomposing them 
into VSH, the first two deformation equations are obtained:
\begin{subequations}\label{eq:disp}
    \begin{eqnarray}
        \frac{\partial \xi_{20}^{(r)}}{\partial r} &=& - q_T f_3 \, \delta T_{20} + f_1 \, \xi_{20}^{(r)} + f_2 \, \xi_{20}^{(\bot)} + f_3 \, t_{20}^{(r)} \\
        \frac{\partial \xi_{20}^{(\bot)}}{\partial r} &=& - \frac{1}{r} \xi_{20}^{(r)} + \frac{1}{r} \xi_{20}^{(\bot)} + \frac{1}{\mu} t_{20}^{(\bot)},
    \end{eqnarray}
\end{subequations}
with
\begin{subequations}
    \begin{eqnarray}
        f_1 &:=& - \left( c_S^2 \frac{\partial \rho}{\partial r} +2 \frac{c_S^2 \rho}{r} - \frac{4}{3} \frac{\mu}{r} \right) f_3 \\
        f_2 &:=& \frac{\lambda}{r} \left( c_S^2 \rho -\frac{2}{3} \mu \right) f_3 \\
        f_3 &:=& \frac{1}{c_S^2 \rho + \frac{4}{3} \mu}.
    \end{eqnarray}
\end{subequations}

On the other hand, the density perturbation can be computed from the continuity equation, which in 
the present case reduces to
\begin{equation}
    \delta \rho = - \vec{\nabla} \cdot \left( \rho \vec{\xi} \right).
\end{equation}
In particular, its $\ell=2$, $m=0$ component 
depends on the temperature perturbation, displacement and traction as
\begin{equation}\label{eq:deltarho}
        \delta \rho =\rho q_T f_3 \, \delta T_{20} - g_1 \, \xi^{(r)}_{20} + \rho g_2 \, \xi_{20}^{(\bot)} - \rho f_3 t_{20}^{(r)},
\end{equation}
with
\begin{subequations}
    \begin{eqnarray}
        g_1 &:=& \frac{\partial \rho}{\partial r} + \frac{2 \rho}{r} + \rho f_1 \\
        g_2 &:=& 2 \lambda \frac{\mu}{r} f_3.
    \end{eqnarray}
\end{subequations}

The expression for the density and pressure perturbations in Eqs.~\ref{eq:deltarho} 
and~\ref{eq:deltap} can be substituted into Eq.~\ref{eq:mom} in order to find the other two 
deformation equations. Those are
\begin{subequations}\label{eq:trac}
    \begin{eqnarray}
        \frac{\partial t^{(r)}_{20}}{\partial r} = - f^{(r)}_{20} + H_T \delta T_{20} + H_1 \xi^{(r)}_{20} + H_2 \xi_{20}^{(\bot)} + H_3 t^{(r)}_{20} + 2 \frac{\lambda}{r} t^{(\bot)}_{20} \\
        \frac{\partial t^{(\bot)}_{20}}{\partial r} = - f_{20}^{(\bot)} + \frac{A_T}{r} \delta T_{20} - \frac{A_2}{r} \xi^{(r)}_{20} + Q_\bot \xi^{(\bot)}_{20} - \frac{f_2}{\lambda} t^{(r)}_{20} - \frac{t^{(\bot)}_{20}}{r},
     \end{eqnarray}
\end{subequations}
with
\begin{subequations}
    \begin{eqnarray}
        H_T &:=& \frac{2}{r} A_T + C_2 q_T f_3 + \frac{\partial \Phi}{\partial r} \rho q_T f_3 \\
        H_1 &:=& -\frac{ \partial C_2}{\partial r} - C_2 f_1 -\frac{2}{r} B_2 - g_1 \frac{\partial\Phi}{\partial r} \\
        H_2 &:=& -2\lambda \frac{\mu}{r^2} - \frac{4 \mu}{r} f_2 - C_2 f_2 + \rho g_2 \frac{\partial \Phi}{\partial r} \\
        H_3 &:=& \left(-\frac{4 \mu}{r} - C_2 - \rho \frac{\partial \Phi}{\partial r}\right) f_3 \\
        A_T &:=& 2 \mu q_T f_3 \\
        A_2 &:=& c_S^2 g_1 - \frac{1}{3} B_2 \\
        Q_\bot &:=& \frac{A_3}{r} - \frac{2\mu}{r^2} (1-\lambda)
    \end{eqnarray}
    \begin{eqnarray}
        C_2 &:=& c_S^2 g_1 + \frac{2}{3} B_2 \\
        B_2 &:=& 2 \mu \left( f_1 -\frac{1}{r} \right).
    \end{eqnarray}
\end{subequations}

Equations~\ref{eq:disp} and~\ref{eq:trac} form together a non-homogeneous linear system of 
differential equations. The 
forcing term of that system includes both the effects of the magnetic field through the 
VSH components $f^{(r)}_{20}$ and $f^{(\bot)}_{20}$ of the Lorentz force density $\vec{f}$ found 
in Sec.~\ref{sec:magn}, and the impact of thermal perturbations, which are plugged in 
from the $\delta T_{20}$ profile computed in Sec.~\ref{sec:therm}.  As in Sec.~\ref{sec:therm}, first the non-homogeneous part is solved 
once for each independent variable, then solved with all variables 
vanishing at one end of the integration interval, and finally the solutions are combined to satisfy the boundary conditions. Once again a Radau IIA method implemented by the 
\verb|solve_ivp| function of the SciPy Python library is used for the numerical solutions.

The system of Eqs.~\ref{eq:disp} and~\ref{eq:trac} is solved in the crust, which 
means between $r=r_{cc}$ and $r=r_S$. 
The boundary conditions on the system imply that all components of the traction are continuous 
at both interfaces. In addition, since both the core and the ocean are considered liquid because 
they lack internal elastic forces and shear stresses, the traction is 
just $\vec{t} = -\delta p \,\hat{r}$. Using the expression of the temperature perturbation in the 
liquid regions of the star found in Sec.~\ref{sec:core}, the boundary conditions are
\begin{subequations}
    \begin{eqnarray}
        t_{20}^{(r)} (r_{cc}) &=& - r_{cc}f_{20}^{(\bot)} (r_{cc}) \\
        t_{20}^{(r)} (r_{S}) &=& - r_{S}f_{20}^{(\bot)} (r_S) \\
        t_{20}^{(\bot)} (r_{cc})&=& t_{20}^{(\bot)} (r_{S}) = 0.
    \end{eqnarray}
\end{subequations}

It can be shown that, in the case of a poloidal only magnetic field, the only non-zero VSH components 
with $\ell=2$ of the Lorentz force density are $f_{20}^{(r)}$ and $f_{20}^{(\bot)}$. Also, 
the only relevant component of the temperature perturbation is $\delta T$. By linearity, 
this justifies the simplifying assumption of considering only the corresponding components of 
the traction and displacement vector fields.

The total entity of the deformation of a NS is usually summarized by its ellipticity $\varepsilon\approx{Q_{22}}/{I}$, where $Q_{22}$ is the quadrupole, i.e. the $l=m=2$ mass multipole, and $I$ the moment of inertia. However, 
the deformation equations presented in this section describe the shape of the star as a 
spheroid, as we consider a non rotating model as a background, and therefore compute an axisymmetric deformation with respect to the magnetic axis. In this case the ellipticity is better
replaced by the eccentricity of the NS's section parallel to the magnetic
axis itself. The eccentricity is defined in~\cite{GWbook} as
\begin{equation}
    e := \frac{I_{33} - I_{11}}{I_{11}},
\end{equation}
where $I_{ij}$ with $i,j = 1, 2, 3$ are the components of the inertia tensor.

Using the definition of the inertia tensor and keeping only the lowest order terms, the eccentricity 
can be rewritten in terms of the density as
\begin{equation}
    e = - \frac{3}{4 \sqrt{5\pi}} \frac{\int_0^R \delta \rho_{20}  r^4 dr}{\int_0^R \rho r^4 dr}.
\end{equation}
Spherical harmonics of $\delta \rho$ with $\ell \neq 2$ are eliminated by the angular integrals in 
the inertia tensor, which justifies the simplifying assumption of defining deformation equations 
only for $\ell = 2$ components.

Note that in a real NS the magnetic axis will be inclined with respect to the rotation axis, so that the deformation computed in this section will acquire also $m=1$ and $m=2$ components with respect to a spherical harmonic basis aligned with said rotation axis. The ellipticity will therefore differ from the eccentricity computed here only by a geometric factor depending on the inclination angle. We therefore take it as a reasonable order of magnitude estimate, as discussed in \cite{haskell08}.

\section{Results}
\label{results}

Solving the equations presented in the previous sections for different input parameters 
leads to computing the eccentricity of the star under those global parameters. 

For each background model, meaning for each combination of values of $\dot{M}$ and $\Delta$ 
described in Sec.~\ref{sec:bgTemp}, the temperature perturbation equations presented in Sec.~\ref{sec:therm} and the deformation equations of Sec.~\ref{sec:defCrust} are 
solved for different values of polar magnetic field $B_p$ and accretion asymmetry $D$. The 
values considered for the magnetic field are $B_p = \SI{1e8}{\gauss}$, $B_p = \SI{1e9}{\gauss}$, 
$B_p = \SI{1e10}{\gauss}$ and $B_p = \SI{1e11}{\gauss}$. The values considered for the 
accretion asymmetry are $D=0$, $D=0.0005$, $D=0.001$, $D=0.002$, $D=0.005$, $D=0.05$, $D=0.1$, 
$D=0.5$ and $D=1$. The range chosen for $D$, except from $D=0$, includes values as low as 
$D\sim \num{1e-3}$, as such accretion asymmetry is found by the simulations of the ocean 
MHD equilibrium presented in~\cite{haskell20}.

However, some combinations of the values listed above describe a regime that falls outside the 
scope of the model. Indeed, in order for the deformation equations described in 
Sec.~\ref{sec:defCrust} to hold, the internal 
forces in the crust must be in the elastic regime. This is quantified through the Von Mises strain, 
which is defined in~\cite{UCB2000} as
\begin{equation}
    \overline{\sigma}^2 := \frac{1}{8\mu^2} \sum_{i,j}\sigma_{ij} \sigma_{ij},
\end{equation}
where $i,j = 1,2,3$. If $\overline{\sigma}>0.1$, 
the lattice either breaks or deforms into a plastic flow and internal forces can no longer 
be described through the strain tensor \citep{horowitz09}. Thus, combinations of input values that produce 
$\overline{\sigma} >0.1$ at any point in the crust, lead to predicted deformations predicted 
that can no longer be considered small. The results produced by such input values 
of the parameters are excluded from this article, as a static model like the one presented here is 
not equipped to deal with the dynamics of crustal breakup and the results would thence risk to 
be inaccurate and unfounded. 


According to the model here presented, 
the crust may break or yield both in case of a too-large 
accretion asymmetry and in case of a too-high magnetic field at the poles. In particular, 
if the polar magnetic field is $B_p \leq \SI{1e10}{\gauss}$, 
the distribution of accreted material is the only cause of the crust breaking and no configuration 
is found to maintain the elastic regime for $D>\num{0.005}$, 
even if the highest stable value of $D$ is often lower, depending on the total accretion. If the 
magnetic field is $B_p =\SI{1e11}{\gauss}$, the same can be said only in the case of a normal 
core, if the total accretion rate is $\dot{M} \geq \SI{6e-9}{\Msun\per\year}$. In other cases 
with the same magnetic field, either the crust breaks for any value od $D$, or there exist a 
slim interval of stability for that parameter (the highest being around $D=0.05$ in the case of 
a normal core with $\dot{M}=\SI{5e-10}{\Msun\per\year}$), below which the crust breaks because of 
the magnetic forces and above which it breaks because of accretion asymmetry.

Thus, in most cases, the accretion asymmetry falls in the range 
$10^{-3}\lesssim D \lesssim 10^{-2}$, 
which, interestingly, is the same range found predicted for the accretion in \cite{haskell20}. 
We caution the reader, however, that the simulations in \cite{haskell20} model only the ocean, 
matching to a purely radial field at the crust interface, and exhibit a complex dependence on 
field strength and accreted mass. Future work should therefore couple the two approaches to 
fully capture the dependence of the ellipticity on accretion range and magnetic field strength.
As another example of the restriction of the parameter space caused by crustal breakup, the 
equations were also solved for $\dot{M} = \SI{1e-8}{\Msun\per\year}$ and $B_p = \SI{1e12}{\gauss}$, 
but in that case, $\overline{\sigma}>0.1$ was obtained in some region of the 
crust for all values of $D$ and $\Delta$. 
For smaller values of $\dot{M}$, $B_p = \SI{1e12}{\gauss}$ was instead excluded because 
it would lead to $x_m > 1$ in some part of the crust and a magnetic field that does not act as a 
perturbation with respect to the temperature profile.

The precise values of the eccentricities predicted by the model in the different cases in which the 
crust does not break are presented in App.~\ref{app:ecc}.

For fixed values of  
$\dot{M}$, $\Delta$ and $B_p$, the eccentricity is expected to be linear in $\delta T_{20}(r_S)$, 
since the equations for the temperature variation are linear and the source term of the deformation 
equations is linear in $\delta T$. For this reason, for each combination of $\dot{M}$, $\Delta$ 
and $B_p$, if the crust is predicted not to break for at least three values of $D$ among 
those considered, a least squares fit is performed of the predicted values of $e$ against the 
relation 
\begin{equation}\label{eq:linear}
    e = m_{fit} \, \frac{\delta T_{20} (r_S)}{T_S} + q_{fit}.
\end{equation}

\begin{table}[b!]
\caption{\label{tab:fit}Best fit parameters}
\centering
\begin{tabular}{ccc}
   \hline\hline
    $\log B_p$ [G]  & $m_{fit} \times \num{1e8}$ & $q_{fit}$ \\
    \noalign{\smallskip}
    \multicolumn{3}{c}{Normal core, $\dot{M}=\SI{1e-8}{\Msun\per\year}$} \\
     \hline
    8 & -3.55254(4) & \num{3.5089(4)e-17} \\
    9 & -2.9122(5) & \num{3.510(4)e-15} \\
    10 & -2.88498(18) & \num{3.4791(7)e-13} \\
    11 & -2.87593(14)& \num{3.47824(14)e-11} \\
    \hline
    \noalign{\smallskip}
    \multicolumn{3}{c}{Superfluid core, $\dot{M}=\SI{1e-8}{\Msun\per\year}$} \\
     \hline
    8 & -8.563(6) & \num{9.280(14)e-17} \\
    9 & -6.205(14) & \num{9.286(5)e-15} \\
    10 & -6.188(4)& \num{9.2865(2)e-13} \\
\hline
    \noalign{\smallskip}
    \multicolumn{3}{c}{Normal core, $\dot{M}=\SI{6e-9}{\Msun\per\year}$} \\
     \hline
    8	& -3.0905(11)&	\num{5.288(4)e-17} \\
    9 & -2.9424(4) & \num{5.295(5)e-15} \\
    10 & -2.94237(11) & \num{5.285(2)e-13} \\
    11 & -2.9411(3) & \num{5.2929(4)e-11} \\
    \hline
    \noalign{\smallskip}
    \multicolumn{3}{c}{Superfluid core, $\dot{M}=\SI{6e-9}{\Msun\per\year}$} \\
     \hline    
    8 & -21.14(2)& \num{4.574(5)e-16} \\
\hline
    \noalign{\smallskip}
    \multicolumn{3}{c}{Normal core, $\dot{M}=\SI{2e-9}{\Msun\per\year}$} \\
     \hline
    8 & -13.425(10) & \num{6.270(2)e-16} \\
    \hline
    \noalign{\smallskip}
    \multicolumn{3}{c}{Superfluid core, $\dot{M}=\SI{2e-9}{\Msun\per\year}$} \\
     \hline
    8 & 3.056(2)& \num{-2.6137(2)e-16} \\
    9 & 3.05593(9) & \num{-2.613(4)e-14} \\
    10 & 3.0560(7) & \num{-2.6132(8)e-12} \\
\hline
    \noalign{\smallskip}
    \multicolumn{3}{c}{Normal core, $\dot{M}=\SI{1e-9}{\Msun\per\year}$} \\
     \hline
    8 & 3.4008(12)& \num{-3.75076(2)e-16} \\
\hline
    \noalign{\smallskip}
    \multicolumn{3}{c}{Superfluid core, $\dot{M}=\SI{1e-9}{\Msun\per\year}$} \\
     \hline    
    8 & 0.70722(19) & \num{-1.55238(8)e-16} \\
    9 & 1.3572(3) & \num{-1.52997(4)e-14} \\
    10 & 1.4544(10) & \num{-1.531(3)e-12} \\
\hline
    \noalign{\smallskip}
    \multicolumn{3}{c}{Normal core, $\dot{M}=\SI{5e-10}{\Msun\per\year}$} \\
     \hline
    8 & 0.8167(16) & \num{-2.0058(3)e-16} \\
    9 & 0.74250(15) & \num{1.9980(12)e-14} \\
    10 & 0.7229(10) & \num{-2.004(3)e-12} \\ 
   \hline
    \end{tabular}
    \tablefoot{Best fit parameters for the linear relation between the eccentricity $e$ and the 
    surface temperature perturbation $\delta T_{20} (r_S)$. In the table, $m$ is the slope of the 
    linear relation, while $q$ is the intercept. The uncertainties on $m$ and $q$ is the error on 
    best fit parameters. The values of $m$ and $q$ are shown for different combinations of values 
    of $\dot{M}$, $\Delta$ and $B_p$.}
\end{table}

\begin{figure*}[ht]
    \centering
    \includegraphics[scale=.5]{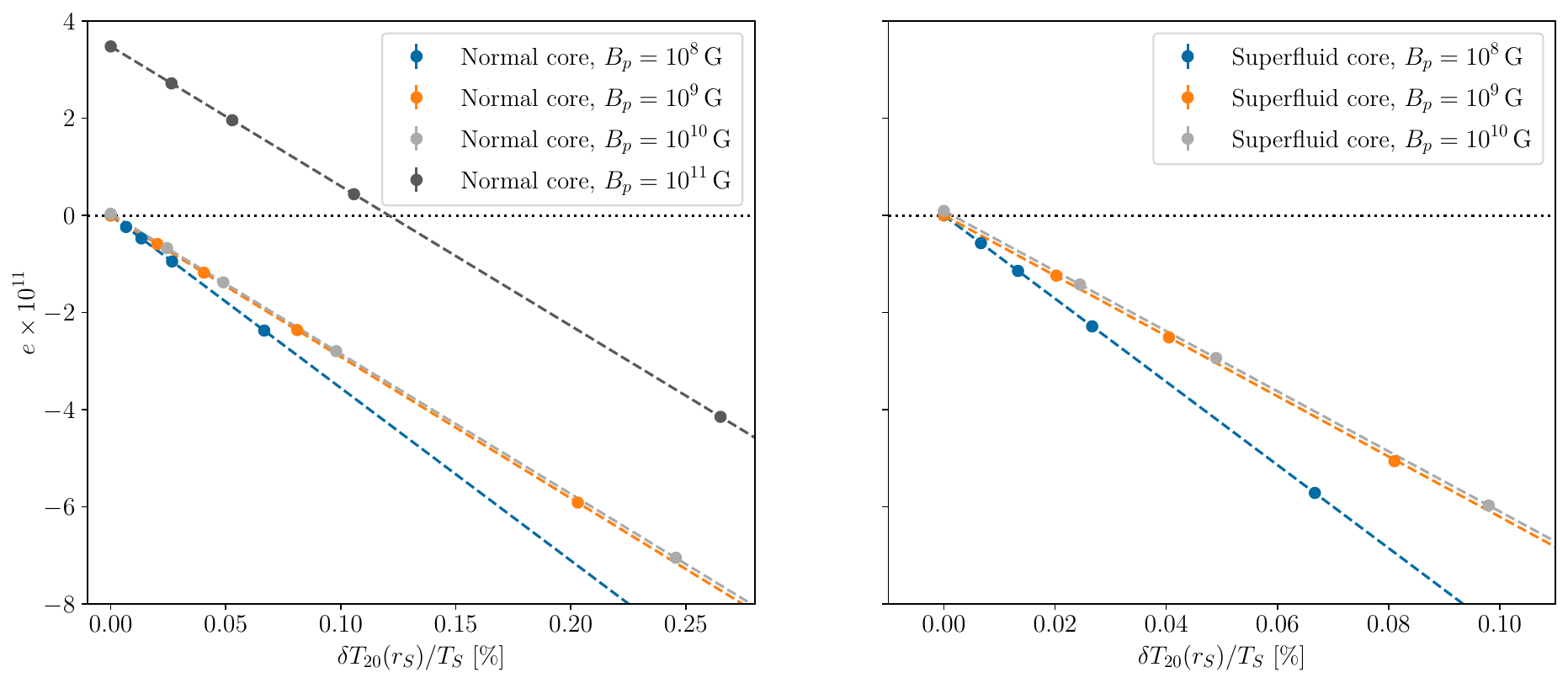}
    {\caption{Eccentricity as a function of the surface temperature deformation in the case 
    of $\dot{M}=\SI{1e-8}{\Msun\per\year}$. The left panel shows the plots for $\Delta=0$, while 
    the right panel shows those for $\Delta =\SI{1}{\mega\eV}$.
    In each panel, the circles represent the computed values of the eccentricities 
    for different values of $\delta T_{20} (r_S)$, hence of $D$, 
    while each dotted line is the best-fit curve  
    for the computed values of the same color. The black dotted line represents $e=0$.}
    \label{fig:fit8}}
\end{figure*}

\begin{figure}[ht]
    \centering
    \includegraphics[width=\hsize]{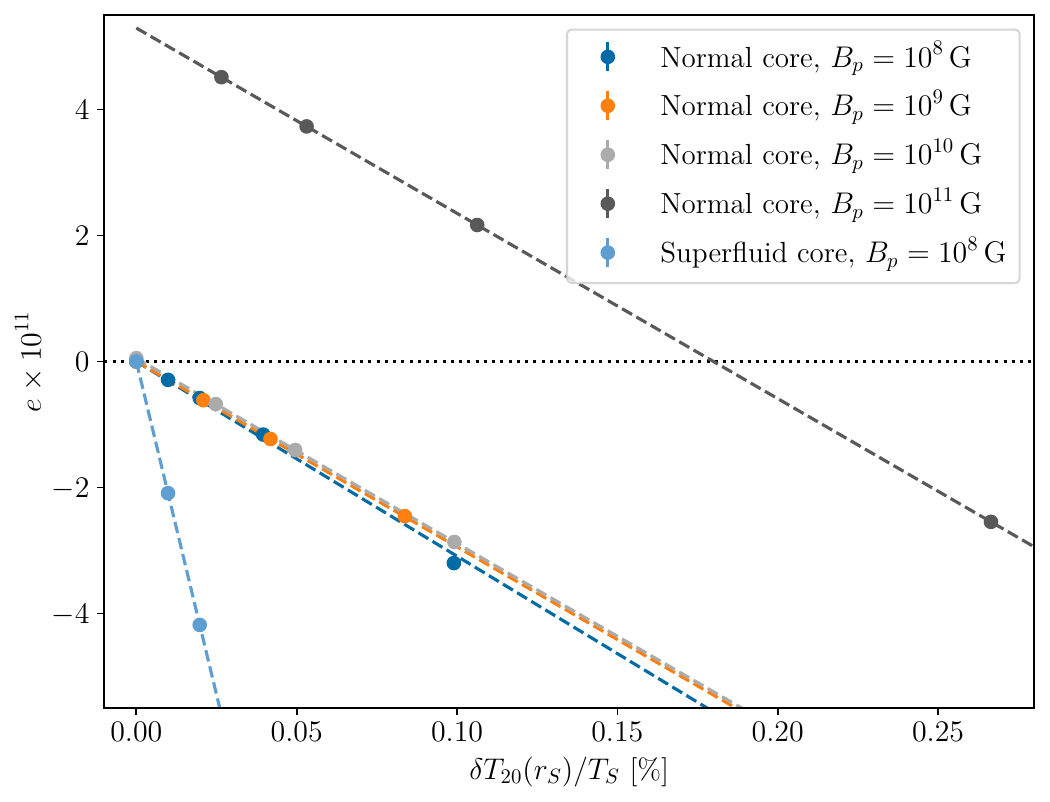}
    \caption{Eccentricity as a function of the surface temperature deformation in the case 
    of $\dot{M}=\SI{6e-9}{\Msun\per\year}$. 
    In each panel, the circles represent the computed values of the eccentricities 
    for different values of $\delta T_{20} (r_S)$, hence of $D$, 
    while each dotted line is the best-fit curve  
    for the computed values of the same color. The black dotted line represents $e=0$.}
    \label{fig:fit6}
\end{figure}

\begin{figure}[ht]
    \centering
    \includegraphics[width=\hsize]{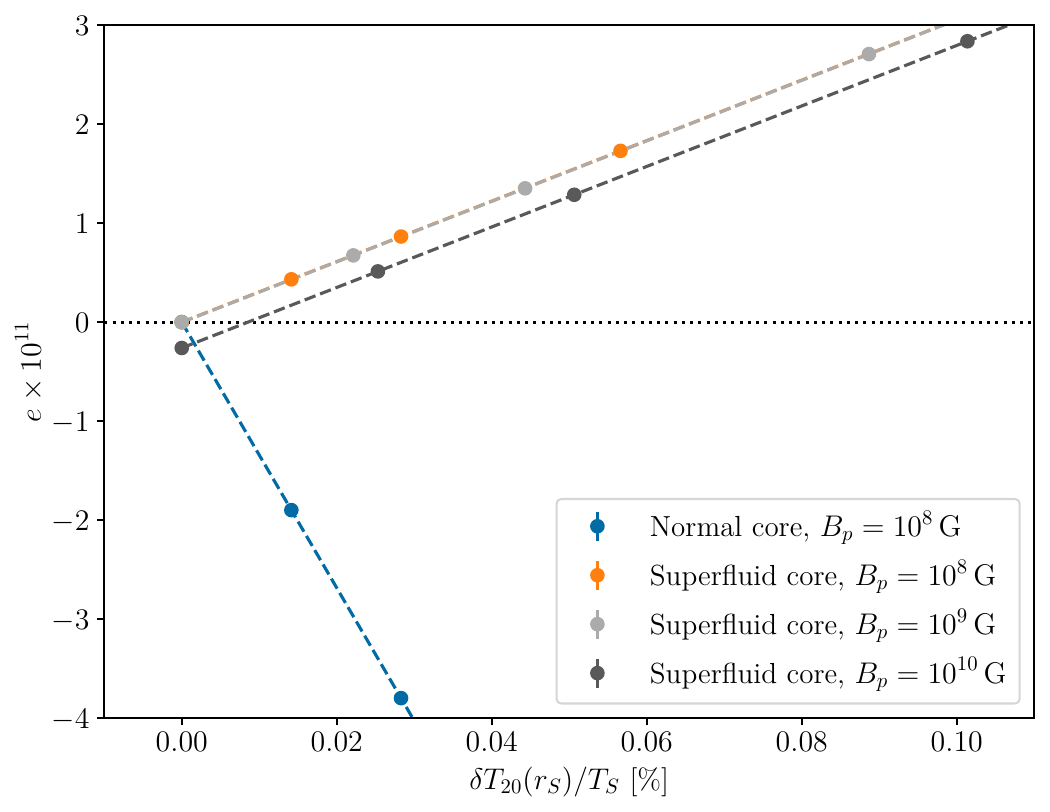}
        \caption{Same as Fig.~\ref{fig:fit6} but for $\dot{M}=\SI{2e-9}{\Msun\per\year}$.}
            \label{fig:fit2}
\end{figure}

\begin{figure}[ht]
    \centering
    \includegraphics[width=\hsize]{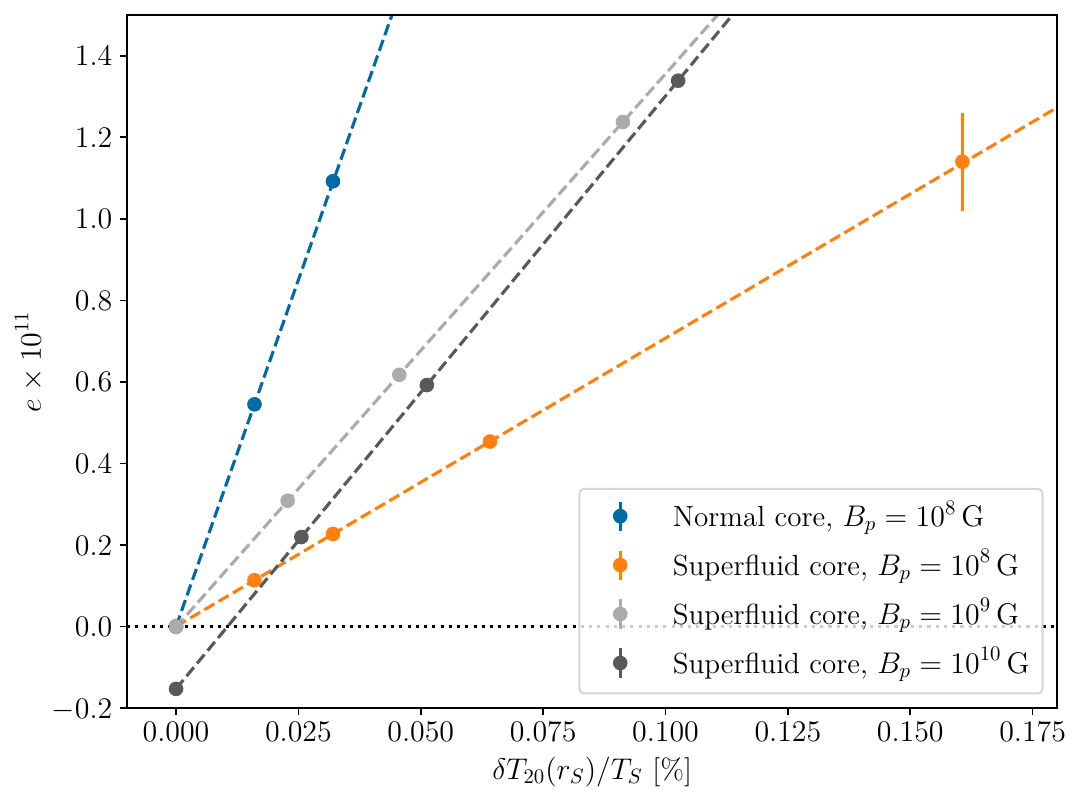}
        \caption{Same as Fig.~\ref{fig:fit6} but for $\dot{M}=\SI{1e-9}{\Msun\per\year}$.}
            \label{fig:fit9}
\end{figure}

\begin{figure}[ht]
    \centering
    \includegraphics[width=\hsize]{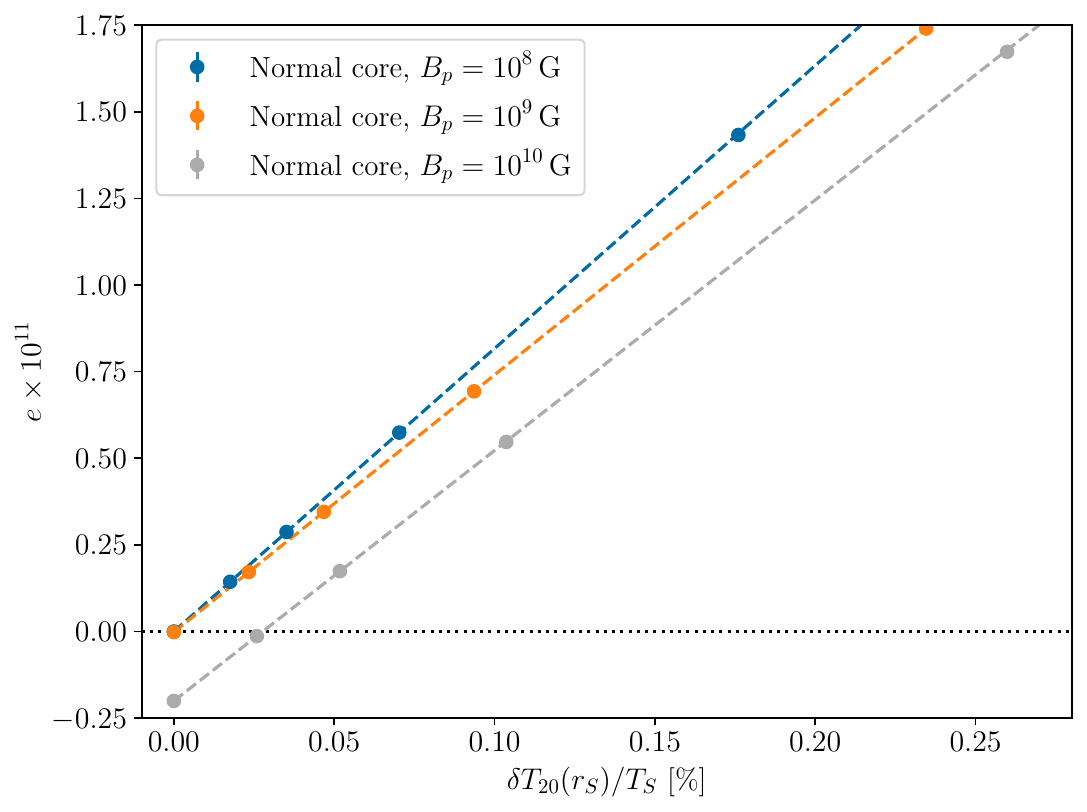}
        \caption{Same as Fig.~\ref{fig:fit6} but for $\dot{M}=\SI{1e-10}{\Msun\per\year}$.}
            \label{fig:fit0}
\end{figure}

The plots showing the predicted eccentricity of the star in different 
conditions as a function of $\delta T_{20} (r_S)$ are presented in Figs.~\ref{fig:fit8} 
through~\ref{fig:fit0}. The best fit values of $m$ and $q$ (Eq.~\ref{eq:linear}) 
are instead presented in Tab.~\ref{tab:fit}.

Those linear fits show that, depending on the total accretion $\dot{M}$, the eccentricity may 
increase or decrease with the accretion asymmetry. In particular, there seems to be a threshold 
total accretion $\dot{M}_{th}$ at which $m_{fit}$ changes sign. The value of $\dot{M}_{th}$ 
depends on the superfluid energy gap in the core, as it can be deduced from figure~\ref{fig:fit2}, 
where, for the same value of $\dot{M}$, $e$ decreases with $\delta T_{20} (r_S)$ with a normal 
core, while it increases with a superfluid core. With $\Delta = 0$, it can be estimated that 
$\SI{1e-9}{\Msun\per\year} \leq \dot{M}_{th} \leq \SI{2e-9}{\Msun\per\year} $, while for 
$\Delta = \SI{1}{\mega\eV}$, the threshold total accretion 
is $\SI{2e-9}{\Msun\per\year} \leq \dot{M}_{th} \leq \SI{6e-9}{\Msun\per\year} $.

The intercept $q_{fit}$ of Eq.~\ref{eq:linear} is always of the opposite sign of $m_{fit}$. 
In addition, an additional least squares fit of the relation between $q_{fit}$ and $B_p$ against 
a power law shows that $q_{fit} \propto B_p^2$. This could be expected since, in the presence of 
a uniform accretion ($D=0$ hence $\delta T_{20} (r_S) =0$), the temperature perturbation only 
depends on the magnetic field 
, which acts through the Lorentz force $\vec{f} \propto B_p^2$ and a source term for temperature perturbations which is proportional to $x_m^2 \propto B_p^2$.

Such linear dependency of the eccentricity on the surface temperature perturbation leads $e$ to 
change sign with an increase of $D$ and shows a competition between the magnetic field and the 
accretion asymmetry. In the case of larger $\dot{M}$, where $m_{fit}<0$, 
at low values of $D =0$ and in the presence of a large magnetic field the eccentricity is positive, 
while if the accretion asymmetry prevails on the Lorentz force $e<0$. This is the expected 
behavior, since a negative eccentricity corresponds to a prolate spheroid, which should be 
produced by accreted matter depositing at the magnetic poles, while a positive $e$ (which means 
an oblate spheroid) is the kind of deformation that is produced in the presence of just a 
poloidal magnetic field (neglecting temperature perturbations) according 
to \cite{2021PASA...38...43S}, \cite{2011MNRAS.417.2696P} and \cite{2009MNRAS.395.2162L}. 

An unexpected result, instead shows up in case of $\dot{M} < \dot{M}_{th}$, where the competition is 
inverted, as large magnetic fields produce prolate spheroids while high values of $D$ are associated 
to oblate stars. A possible explanation for this inverted behavior is that, in general, the deformation of a neutron star are dominated by the temperature perturbations rather than the Lorentz force. As a consequence, the indirect effect of the magnetic field on the deformations through the source term of temperature perturbation equations~\ref{eq:deltaT} and~\ref{eq:deltaF} may be more 
relevant than the direct effect of the Lorentz force. Indeed, according to the model here presented, a $B_p = \SI{1e10}{\gauss}$ magnetic field in a star with a superfluid core and a total accretion $\dot{M}=\SI{1e-9}{\Msun\per\year}$ would show an inverted behavior and, with symmetrical accretion, 
it would have a maximum temperature variation $\max \left| \frac{\delta T_{20}}{T}\right| \sim \num{5e-5}$, which by itself, according to~\cite{UCB2000}, would account for an eccentricity of about $|e| \sim \num{2.5e-10}$. On the other hand, the same poloidal magnetic field, neglecting the rigidity of the crust and reaction layers, would produce an eccentricity of $|e| \sim \num{2e-14}$, according to~\cite{haskell08}. However, we are not able to provide simple and convincing interpretation for this phenomenon at the moment.


\begin{figure}[t!]
    \centering
    \includegraphics[width=\hsize]{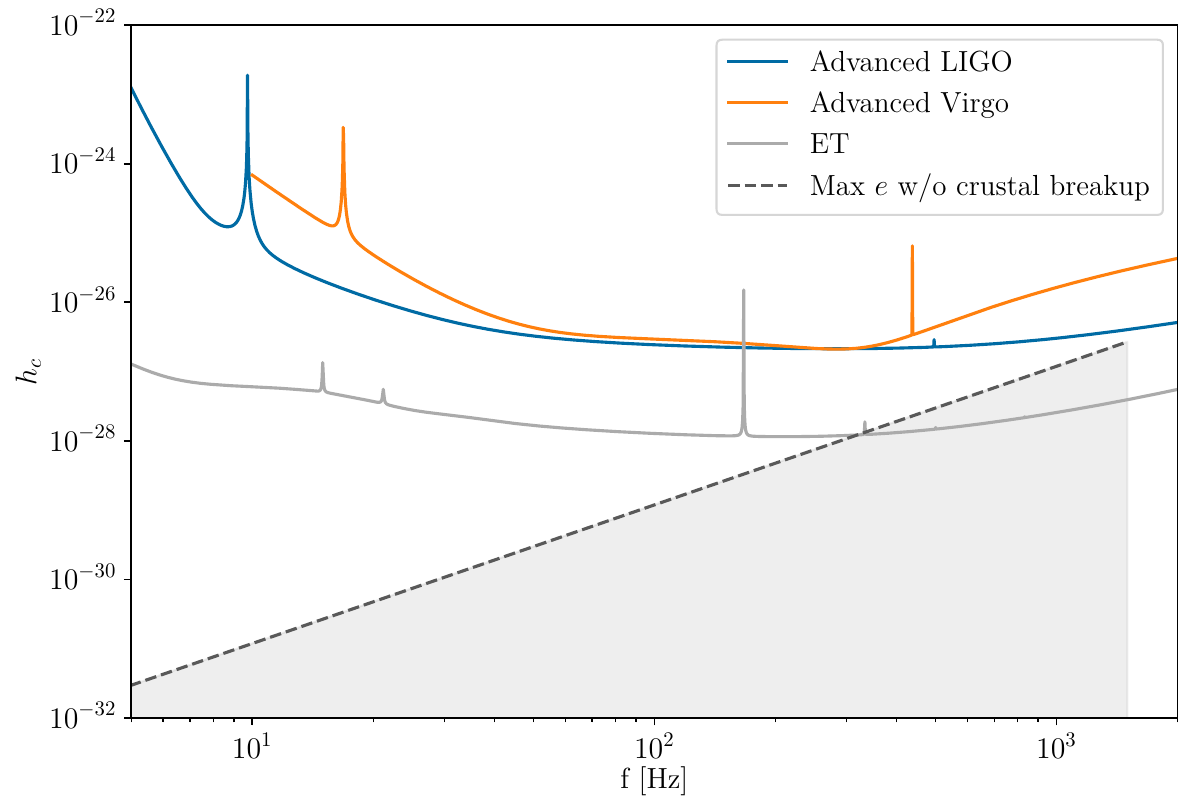}
    \caption{Comparison between the predicted GW signal of deformations produced by the model 
    here presented and the sensitivities of the current LIGO and Virgo interferometers and of 
    the planned Einstein Telescope (ET). The maximum predicted characteristic strain is represented 
    as a function of the rotation frequency $f$ of the star by the dotted line. That signal 
    is produced by the maximum eccentricity computed within the model here presented without 
    breaking the crust of the NS. It corresponds to $e=\num{-7.046(9)e-11}$, which is realized in 
    the case of $\dot{M} = \SI{1e-8}{\Msun\per\year}$, $\Delta =0$, 
    $B_p=\SI{1e10}{\gauss}$ and $D=0.005$. Signals produced according to the model here presented 
    by known pulsars should lie in the shaded area below the dashed line. The sensitivity curves 
    of the GW detectors, represented by solid lines, are taken from \cite{LVsens} for LIGO and 
    Virgo and from  \cite{ETsens} for ET.}
    \label{fig:detectors}
\end{figure}

For what concerns the observational prospects of the model here presented, the main way to 
measure the predicted eccentricities would be the detection of GWs with 
next-generation interferometers. Indeed, a spheroidal neutron star which rotates around an axis 
other than its axis of symmetry with rotation frequency $\omega$, according 
to \cite{1979PhRvD..20..351Z}, would emit GWs with frequencies $\omega$ and $2\omega$ and with 
maximum characteristic strain
\begin{equation}\label{eq:strain}
    h_c = \frac{4 \pi^2 G}{c^4 d} I_0\, |e| \, \omega^2,
\end{equation}
where $d$ is the distance from the source and $I_0 := \int_0^R \rho \, r^4 dr $ is the 
unperturbed moment of inertia of the star. The actual measured value of the characteristic 
strain could be less than $h_c$ since that value is multiplied by a factor of order unity which 
depends on the angle between the magnetic axis and the rotation axis and on the angle between the 
latter and the line of sight. Equation~\ref{eq:strain} represents the observed characteristic strain 
in case the magnetic axis is perpendicular to the rotation ais, which is in turn aligned to the 
line of sight. As a consequence, the GW signals produced by known pulsars due to deformations 
predicted by the model should lie in the shaded area of Fig.~\ref{fig:detectors}, which in addition compares the predictions to sensitivities current and future interferometers.

The sign of the eccentricity is relevant for observational prospects of the model. Indeed, 
according to \cite{2002PhRvD..66h4025C}, the free precession of a spheroidal NS around a rotation 
axis that does not coincide with its axis of symmetry is in general unstable. In particular, the 
magnetic axis of an oblate star tends to align with the rotation axis, while a prolate star ends up 
with the axis of symmetry perpendicular to the rotation axis after a transient. This means that 
$e<0$ realizes the optimal condition for the observation of GWs from the deformed NS.

\section{Conclusions}
\label{conclu}

We have presented a model for the deformation of accreting magnetized neutron stars that combines, within a single framework, the effects of magnetic stresses, asymmetric crustal heating due to accretion and the elastic response of the crust. By solving the thermal evolution and the linearised deformation equations consistently, we computed the equilibrium quadrupolar deformation produced for a range of magnetic fields, accretion rates and core superfluid properties.

Our results show that the magnetic field and the thermally-induced deformation associated with asymmetric accretion generally compete with one another. We identify a threshold accretion rate $\dot{M}_{th}\approx 0.1 \dot{M}_{Edd}$, whose exact value depends on the assumed superfluid energy gap, above which magnetic stresses favor oblate configurations while asymmetric accretion produces prolate deformations. Below this threshold the behavior is reversed. Although we do not yet have a simple physical explanation for this inversion, its robustness across our models suggests that it reflects a genuine feature of the coupled magneto-thermal response. 
This suggests the presence of a complicated interplay between the magnetic field and the temperature perturbations where the indirect effect of the first through the latter is relevent.

We caution however that the deformation depends not only on the magnetic field strength but also on the efficiency with which asymmetric accretion is communicated through the ocean to the crust, encapsulated here by the parameter D. Determining this parameter from first principles will require coupling the present crustal calculation to magnetohydrodynamic models of the ocean and accretion flow. Precise modelling of the crust-ocean interface may therefore shed light on the nature of the solutions below and above $\dot{M}_{th}$ and will be the focus of future work.

The predicted eccentricities reach values of $e\approx 10^{-11}$
 without exceeding the crustal breaking strain, implying characteristic continuous gravitational-wave amplitudes that are generally below the sensitivity of current interferometers but could become accessible to third-generation detectors such as the Einstein Telescope or Cosmic Explorer. These results are therefore consistent with the current non-detection of continuous gravitational waves from accreting neutron stars while providing quantitative predictions for future searches.

Finally, the present model also provides a flexible framework within which more realistic models for the crust or the magnetic field may be incorporated, such as e.g. twisted-torus models for the magnetic field \citep{2021PASA...38...43S} or equations of state that include pasta phases at the base of the crust \citep{2025ApJ...980..144L}.

More generally, our results demonstrate that magnetic stresses, thermal asymmetries and crustal elasticity cannot always be considered independently when modelling neutron-star mountains. As continuous gravitational-wave searches become increasingly sensitive, models incorporating these coupled effects will be important both for interpreting future detections and for placing meaningful constraints on the internal physics of neutron stars.



\section*{Acknowledgments}

This 
publication is based upon work from COST Action SCALES CA24139, supported by COST (European Cooperation in Science and Technology).
This work was partially supported by the Polish National Science Centre grants No. 2023/49/B/ST9/02777 and 2021/43/B/ST9/01714. 
The authors also thank D.I. Jones a P. Covas for the useful comments during the preliminary circulation of the article.

\clearpage 
\newpage 
\appendix
\nolinenumbers

\section{Predicted Eccentricities}
\label{app:ecc}

The model here presented allows for a precise determination of the eccentricity of a NS, 
starting from a selection of parameters, through the numerical solution of the deformation 
equations described in Sec.~\ref{sec:defCrust}. Tables~\ref{tab:8n} to~\ref{tab:0n} gather the 
eccentricities predicted for the vales of superfluid energy gap in the core, total accretion rate, 
polar magnetic field and accretion asymmetry such that the temperature variation is perturbative 
at all point in the crust, the forces inside the crust remain elastic and the crust does not break 
or yield. There is a separate table for each combination considered of values of the 
superfluid energy gap and the total accretion rate. In each table, the uncertainty on the eccentricity corresponds to the upper bound of the global truncation error of the numerical method used in that instance to solve the deformation equations. 

Tables~\ref{tab:8n} through~\ref{tab:0n} also include the values of the relative temperature perturbation at the neutron drip point $\left.\frac{\delta T}{T}\right|_{r=r_{nd}} $ corresponding to each predicted value of the eccentricity. This data is useful to show that the sign of the eccentricity correspond to the sign of the temperature perturbation at neutron drip for all cases considered, suggesting that reaction in that region dominate the temperature perturbations and thence the deformation, as those are the electron captures that release the most energy per accreted nucleon, according to~\cite{EoS18}. It also quantifies the impact of the magnetic source term in temperature perturbation equations even in the case of symmetric accretion. The uncertainty on $\delta T (r_{nd})$ corresponds to the error of the numerical solution of equations~\ref{eq:deltaT} and~\ref{eq:deltaF}.

\begin{table}[ht]
\centering
 \caption{\label{tab:8n}Normal core, $\dot{M} = \SI{1e-8}{\Msun\per\year}$}
 \begin{tabular}{ccc}
   \hline\hline
    $D$ & $e$ & $\left.\frac{\delta T}{T}\right|_{r=r_{nd}} $ \\
    \noalign{\smallskip}
    \multicolumn{3}{c}{$B_p=\SI{1e8}{\gauss}$} \\
     \hline
    0 & \num{3.5089(4)e-17} & \num{7.5796(8)E-13} \\
    0.0005 & \num{-2.3694(3)e-12} & \num{-3.7510(4)E-08} \\
    0.001 & \num{-4.736(7)e-12} & \num{-7.5031(8)E-08}\\
    0.002 & \num{-9.475(3)e-12} & \num{-1.5011(2)E-07}\\
    0.005 & \num{-2.3707(3)e-11} & \num{-3.7560(4)E-07}\\
    \hline
    \noalign{\smallskip}
    \multicolumn{3}{c}{$B_p=\SI{1e9}{\gauss}$} \\
    \hline
    0 & \num{3.510(4)e-15} & \num{7.7079(8)E-11} \\
    0.0005 & \num{-5.89(2)e-12} & \num{-9.2926(9)E-08} \\
    0.001 & \num{-1.1789(2)e-11} & \num{-1.8600(2)E-07} \\
    0.002 & \num{-2.3597(16)e-11} & \num{-3.7234(4)E-07} \\
    0.005 & \num{-5.912(5)e-11} & \num{-9.3299(9)E-07} \\
    \hline
    \noalign{\smallskip}
    \multicolumn{3}{c}{$B_p=\SI{1e10}{\gauss}$} \\
    \hline
    0 & \num{3.4791(7)e-13} & \num{7.4910(7)E-09} \\
    0.0005 & \num{-6.7122(2)e-12} & \num{-1.03559(10)E-07} \\
    0.001 & \num{-1.3778(3)e-11} & \num{-2.1470(2)E-07} \\
    0.002 & \num{-2.7928(5)e-11} & \num{-4.3729(4)E-07} \\
    0.005 & \num{-7.046(9)e-11} & \num{-1.10734(11)E-06} \\
    \hline
    \noalign{\smallskip}
    \multicolumn{3}{c}{$B_p=\SI{1e11}{\gauss}$} \\
    \hline
    0 & \num{3.4783(4)e-11} & \num{7.4906(7)E-07} \\
    0.0005 & \num{2.7190(3)e-11} & \num{6.2922(6)E-07} \\
    0.001 & \num{1.95911(8)e-11} & \num{5.0927(5)E-07} \\
    0.002 & \num{4.3716(4)e-12} & \num{2.6903(3)E-07} \\
    0.005 & \num{-4.1457(9)e-11} & \num{-4.5436(5)E-07} \\
   \hline
    \end{tabular}
\end{table}

\begin{table}[ht]
\centering
 \caption{\label{tab:8s}Superfluid core, $\dot{M} = \SI{1e-8}{\Msun\per\year}$}
 \begin{tabular}{ccc}
   \hline\hline
    $D$ & $e$ & $\left.\frac{\delta T}{T}\right|_{r=r_{nd}} $ \\
        \noalign{\smallskip}
    \multicolumn{3}{c}{$B_p=\SI{1e8}{\gauss}$} \\
     \hline
    0 & \num{9.280(14)e-17} & \num{1.8476(2)E-11} \\
    0.0005 & \num{-5.71(3)e-12} & \num{-1.00799(10)E-06} \\
    0.001 & \num{-1.142(3)e-11} & \num{-2.0163(2)E-06} \\
    0.002 & \num{-2.284(2)e-11} & \num{-4.0338(4)E-06} \\
    0.005 & \num{-5.713(11)e-11} & \num{-1.00933(10)E-05} \\
    \hline
        \noalign{\smallskip}
    \multicolumn{3}{c}{$B_p=\SI{1e9}{\gauss}$} \\
     \hline
    0 & \num{9.286(5)e-15} & \num{1.8470(2)E-09} \\
    0.0005 & \num{-1.24(2)e-11} & \num{-2.5055(3)E-06} \\
    0.001 & \num{-2.511(6)e-11} & \num{-5.0146(5)E-06} \\
    0.002 & \num{-5.06(4)e-11} & \num{-1.00384(10)E-05} \\
    \hline
        \noalign{\smallskip}
    \multicolumn{3}{c}{$B_p=\SI{1e10}{\gauss}$} \\
    \hline
    0 & \num{9.2865(2)e-13} & \num{1.8472(2)E-07} \\
    0.0005 & \num{-1.422(2)e-11} & \num{-2.8182(3)E-06} \\
    0.001 & \num{-2.937(5)e-11} & \num{-5.8237(6)E-06} \\
    0.002 & \num{-5.971(5)e-11} & \num{-1.18420(11)E-05} \\
    \hline
        \noalign{\smallskip}
    \multicolumn{3}{c}{$B_p=\SI{1e11}{\gauss}$} \\
    \hline
    0.002 & \num{2.76097(4)e-11} & \num{5.5317(5)E-06} \\
   \hline
    \end{tabular}
\end{table}

\begin{table}[ht]
\centering
 \caption{\label{tab:6n}Normal core, $\dot{M} = \SI{6e-9}{\Msun\per\year}$}
 \begin{tabular}{ccc}
   \hline\hline
    $D$ & $e$ & $\left.\frac{\delta T}{T}\right|_{r=r_{nd}} $ \\
        \noalign{\smallskip}
    \multicolumn{3}{c}{$B_p=\SI{1e8}{\gauss}$} \\
     \hline
    0 & \num{5.288(4)e-17} & \num{1.07861(11)E-11} \\
    0.0005 & \num{-2.913(6)e-12} & \num{-5.8596(6)E-07} \\
    0.001 & \num{-5.802(15)e-12} & \num{-1.17214(12)E-06} \\
    0.002 & \num{-1.16(2)e-11} & \num{-2.3451(2)E-06} \\
    0.005 & \num{-3.200(2)e-11} & \num{-5.8689(6)E-06} \\
    \hline
        \noalign{\smallskip}
    \multicolumn{3}{c}{$B_p=\SI{1e9}{\gauss}$} \\
     \hline
    0 & \num{5.295(5)e-15} & \num{1.07667(11)E-09} \\
    0.0005 & \num{-6.1474(11)e-12} & \num{-1.23648(12)E-06} \\
    0.001 & \num{-1.2305(2)e-11} & \num{-2.4749(2)E-06} \\
    0.002 & \num{-2.455(6)e-11} & \num{-4.9546(5)E-06} \\
    \hline
        \noalign{\smallskip}
    \multicolumn{3}{c}{$B_p=\SI{1e10}{\gauss}$} \\
     \hline
    0 & \num{5.285(2)e-13} & \num{1.07434(10)E-07} \\
    0.0005 & \num{-6.7575(2)e-12} & \num{-1.35736(14)E-06} \\
    0.001 & \num{-1.4050(6)e-11} & \num{-2.8234(3)E-06} \\
    0.002 & \num{-2.8654(7)e-11} & \num{-5.7592(6)E-06} \\
    \hline
        \noalign{\smallskip}
    \multicolumn{3}{c}{$B_p=\SI{1e11}{\gauss}$} \\
     \hline
    0.0005 & \num{4.513(4)e-11} & \num{9.1892(9)E-06} \\
    0.001 & \num{3.7313(3)e-11} & \num{7.6181(8)E-06} \\
    0.002 & \num{2.1662(3)e-11} & \num{4.4714(4)E-06} \\
    0.005 & \num{-2.5451(5)e-11} & \num{-5.0038(5)E-06} \\
   \hline
    \end{tabular}
\end{table}

\begin{table}[ht]
\centering
 \caption{\label{tab:6s}Superfluid core, $\dot{M} = \SI{6e-9}{\Msun\per\year}$}
 \begin{tabular}{ccc}
   \hline\hline
    $D$ & $e$ & $\left.\frac{\delta T}{T}\right|_{r=r_{nd}} $ \\
        \noalign{\smallskip}
    \multicolumn{3}{c}{$B_p=\SI{1e8}{\gauss}$} \\
     \hline
    0 & \num{4.574(5)e-16} & \num{1.13633(11)E-10} \\
    0.0005 & \num{-2.090(7)e-11} & \num{-4.5150(5)E-06} \\
    0.001 & \num{-4.181(5)e-11} & \num{-9.0317(9)E-06} \\
    \hline
        \noalign{\smallskip}
    \multicolumn{3}{c}{$B_p=\SI{1e9}{\gauss}$} \\
     \hline
    0 & \num{4.57434(7)e-14} & \num{1.13651(11)E-08} \\
    0.0005 & \num{-3.841(2)e-11} & \num{-9.5389(10)E-06} \\
    \hline
        \noalign{\smallskip}
    \multicolumn{3}{c}{$B_p=\SI{1e10}{\gauss}$} \\
     \hline
    0 & \num{4.570(11)e-12} & \num{1.13654(11)E-06} \\
    0.0005 & \num{-4.094(3)e-11} & \num{-1.01728(10)E-05} \\
    \hline
        \noalign{\smallskip}
    \multicolumn{3}{c}{$B_p=\SI{1e11}{\gauss}$} \\
     \hline
    0.005 & \num{-3.27(5)e-11} & \num{-8.0576(8)E-06} \\
   \hline
    \end{tabular}
\end{table}

\begin{table}[ht]
\centering
 \caption{\label{tab:2n}Normal core, $\dot{M} = \SI{2e-9}{\Msun\per\year}$}
 \begin{tabular}{ccc}
   \hline\hline
    $D$ & $e$ & $\left.\frac{\delta T}{T}\right|_{r=r_{nd}} $ \\
        \noalign{\smallskip}
    \multicolumn{3}{c}{$B_p=\SI{1e8}{\gauss}$} \\
     \hline
    0 & \num{6.270(2)e-16} & \num{1.8390(2)E-10} \\
    0.0005 & \num{-1.899(4)e-11} & \num{-5.3411(5)E-06} \\
    0.001 & \num{-3.798(3)e-11} & \num{-1.06851(10)E-05} \\
    \hline
        \noalign{\smallskip}
    \multicolumn{3}{c}{$B_p=\SI{1e9}{\gauss}$} \\
     \hline
    0 & \num{6.2650(3)e-14} & \num{1.8374(2)E-08} \\
    0.0005 & \num{-2.842(7)e-11} & \num{-8.3439(8)E-06} \\
    \hline
        \noalign{\smallskip}
    \multicolumn{3}{c}{$B_p=\SI{1e10}{\gauss}$} \\
     \hline
    0 & \num{6.265(7)e-12} & \num{1.8374(2)E-06} \\ 
    0.0005 & \num{-2.621(9)e-11} & \num{-7.7215(8)E-06} \\
   \hline
    \end{tabular}
\end{table}

\begin{table}[ht]
\centering
 \caption{\label{tab:2s}Superfluid core, $\dot{M} = \SI{2e-9}{\Msun\per\year}$}
 \begin{tabular}{ccc}
   \hline\hline
    $D$ & $e$ & $\left.\frac{\delta T}{T}\right|_{r=r_{nd}} $ \\
        \noalign{\smallskip}
    \multicolumn{3}{c}{$B_p=\SI{1e8}{\gauss}$} \\
     \hline
    0 & \num{-2.6137(2)e-16} & \num{-1.04073(10)E-10} \\
    0.0005 & \num{4.322(6)e-12} & \num{1.7229(2)E-06} \\
    0.001 & \num{8.645(9)e-12} & \num{3.4467(3)E-06} \\
    0.002 & \num{1.730(2)e-11} & \num{6.8970(7)E-06} \\
    \hline
        \noalign{\smallskip}
    \multicolumn{3}{c}{$B_p=\SI{1e9}{\gauss}$} \\
     \hline
    0 & \num{-2.613(4)e-14} & \num{-1.04046(10)E-08} \\
    0.0005 & \num{6.740(6)e-12} & \num{2.6871(3)E-06} \\
    0.001 & \num{1.35119(4)e-11} & \num{5.3867(5)E-06} \\
    0.002 & \num{2.7072(11)e-11} & \num{1.07923(10)E-05} \\
    \hline
        \noalign{\smallskip}
    \multicolumn{3}{c}{$B_p=\SI{1e10}{\gauss}$} \\
     \hline
    0 & \num{-2.6132(8)e-12} & \num{-1.04056(10)E-06} \\
    0.0005 & \num{5.122(5)e-12} & \num{2.0429(2)E-06} \\
    0.001 & \num{1.2863(5)e-11} & \num{5.1292(5)E-06} \\
    0.002 & \num{2.837(2)e-11} & \num{1.13099(11)E-05} \\
   \hline
    \end{tabular}
\end{table}

\begin{table}[ht]
\centering
 \caption{\label{tab:9n}Normal core, $\dot{M} = \SI{1e-9}{\Msun\per\year}$}
 \begin{tabular}{ccc}
   \hline\hline
    $D$ & $e$ & $\left.\frac{\delta T}{T}\right|_{r=r_{nd}} $ \\
        \noalign{\smallskip}
    \multicolumn{3}{c}{$B_p=\SI{1e8}{\gauss}$} \\
     \hline
    0 & \num{-3.75076(2)e-16} & \num{-4.2559(4)E-10} \\
    0.0005 & \num{5.449(2)e-12} & \num{6.2144(6)E-06} \\
    0.001 & \num{1.092(11)e-11} & \num{1.24308(12)E-05} \\
    \hline
        \noalign{\smallskip}
    \multicolumn{3}{c}{$B_p=\SI{1e9}{\gauss}$} \\
     \hline
    0 & \num{-3.726(16)e-14} & \num{-4.2838(4)E-08} \\
    0.0005 & \num{1.490(12)e-11} & \num{1.6923(2)E-05} \\
    \hline
        \noalign{\smallskip}
    \multicolumn{3}{c}{$B_p=\SI{1e10}{\gauss}$} \\
     \hline
    0 & \num{-3.77541(6)e-12} & \num{-4.2839(4)E-06} \\
    0.0005 & \num{1.420(15)e-11} & \num{1.6135(2)E-05} \\
   \hline
    \end{tabular}
\end{table}

\begin{table}[ht]
\centering
 \caption{\label{tab:9s}Superfluid core, $\dot{M} = \SI{1e-9}{\Msun\per\year}$}
 \begin{tabular}{ccc}
   \hline\hline
    $D$ & $e$ & $\left.\frac{\delta T}{T}\right|_{r=r_{nd}} $ \\
        \noalign{\smallskip}
    \multicolumn{3}{c}{$B_p=\SI{1e8}{\gauss}$} \\
     \hline
    0 & \num{-1.55238(8)e-16} & \num{-1.11307(11)E-10} \\
    0.0005 & \num{1.134(5)e-12} & \num{8.1464(8)E-07} \\
    0.001 & \num{2.268(2)e-12} & \num{1.6296(2)E-06} \\
    0.002 & \num{4.5365(13)e-12} & \num{3.2602(3)E-06} \\
    0.005 & \num{1.14(12)e-11} & \num{8.1570(8)E-06} \\
    \hline
        \noalign{\smallskip}
    \multicolumn{3}{c}{$B_p=\SI{1e9}{\gauss}$} \\
     \hline
    0 & \num{-1.52997(4)e-14} & \num{-1.09751(10)E-08} \\ 
    0.0005 & \num{3.083(13)e-12} & \num{2.2132(2)E-06} \\
    0.001 & \num{6.174(16)e-12} & \num{4.4391(4)E-06} \\
    0.002 & \num{1.2375(3)e-11} & \num{8.8955(9)E-06} \\
    \hline
        \noalign{\smallskip}
    \multicolumn{3}{c}{$B_p=\SI{1e10}{\gauss}$} \\
     \hline
    0 & \num{-1.531(3)e-12} & \num{-1.09741(11)E-06} \\
    0.0005 & \num{2.194(3)e-12} & \num{1.5795(2)E-06} \\
    0.001 & \num{5.922(8)e-12} & \num{4.2588(4)E-06} \\
    0.002 & \num{1.3388(14)e-11} & \num{9.6243(10)E-06} \\
   \hline
    \end{tabular}
\end{table}

\begin{table}[ht]
\centering
 \caption{\label{tab:0n}Normal core, $\dot{M} = \SI{5e-10}{\Msun\per\year}$}
 \begin{tabular}{ccc}
   \hline\hline
    $D$ & $e$ & $\left.\frac{\delta T}{T}\right|_{r=r_{nd}} $ \\
        \noalign{\smallskip}
    \multicolumn{3}{c}{$B_p=\SI{1e8}{\gauss}$} \\
     \hline
    0 & \num{-2.0058(3)e-16} & \num{-2.4046(2)E-10} \\
    0.0005 & \num{1.435(6)e-12} & \num{1.7226(2)E-06} \\
    0.001 & \num{2.871(11)e-12} & \num{3.4467(3)E-06} \\
    0.002 & \num{5.74(2)e-12} & \num{6.8986(7)E-06} \\
    0.005 & \num{1.433(7)e-11} & \num{1.7284(2)E-05} \\
    \hline
        \noalign{\smallskip}
    \multicolumn{3}{c}{$B_p=\SI{1e9}{\gauss}$} \\
     \hline
    0 & \num{-1.9980(12)e-14} & \num{-2.4043(2)E-08} \\
    0.0005 & \num{1.7154(5)e-12} & \num{2.0592(2)E-06} \\
    0.001 & \num{3.4525(11)e-12} & \num{4.1442(4)E-06} \\
    0.002 & \num{6.931(7)e-12} & \num{8.3197(8)E-06} \\
    0.005 & \num{1.740(8)e-11} & \num{2.0889(2)E-05} \\
    \hline
        \noalign{\smallskip}
    \multicolumn{3}{c}{$B_p=\SI{1e10}{\gauss}$} \\
     \hline
    0 & \num{-2.005(9)e-12} & \num{-2.4039(2)E-06} \\
    0.0005 & \num{-1.329(2)e-13} & \num{-1.5684(2)E-07} \\
    0.001 & \num{1.741(3)e-12} & \num{2.0923(2)E-06} \\
    0.002 & \num{5.47(3)e-12} & \num{6.5970(7)E-06} \\
    0.005 & \num{1.673(6)e-11} & \num{2.0161(2)E-05} \\
    \hline
        \noalign{\smallskip}
    \multicolumn{3}{c}{$B_p=\SI{1e10}{\gauss}$} \\
     \hline
    0.05 & \num{2.614(4)e-12} & \num{3.4057(3)E-06} \\
   \hline
    \end{tabular}
\end{table}

\end{document}